\documentclass[aps,prb,twocolumn]{revtex4}
\usepackage{graphicx}
\usepackage{amsmath}
\usepackage{amssymb}
\usepackage{braket}
\usepackage{hyperref}
\usepackage{subfigure}
\usepackage[usenames,dvipsnames]{color}

\usepackage{graphicx}
\usepackage{subfigure}

\newcommand{\ct}{\cite}

\newcommand{\bi}{\bibitem}
\newcommand{\be}{\begin{equation}}
\newcommand{\ee}{\end{equation}}
\newcommand{\ba}{\begin{eqnarray}}
\newcommand{\ea}{\end{eqnarray}}

\begin{document}

%\title{Reentrance of areas of FZs and dynamical phase transitions  in the  linearly ramped topological Haldane model}
%\title{Dynamical quantum phase transitions and  connection to equilibrium topology: linearly ramped Haldane model}
\title{The fate of current, residual energy and entanglement entropy in aperiodic driving  of one dimensional Jordan Wigner integrable models}
\author{Somnath Maity, Utso Bhattacharya and Amit Dutta \\
Department of Physics, Indian Institute of Technology, 208016, Kanpur}

\begin{abstract}

We investigate  the dynamics of two Jordan Wigner solvable models, namely, the one dimensional chain of hard-core bosons (HCB)  and the  one-dimensional transverse field Ising model
under  coin-toss like  aperiodically driven   staggered on-site potential and the transverse field, respectively. It is demonstrated that both the models heat up to the infinite
temperature ensemble for a minimal aperiodicity in driving.  Consequently, in the case of the HCB chain, we show that the initial current generated by the application of a twist vanishes in the asymptotic
limit for any driving frequency. For the transverse Ising chain, we establish  that the system not only reaches the  diagonal ensemble but  the entanglement also attains the thermal value in the asymptotic limit following initial ballistic growth. All these
findings, contrasted with that of the perfectly periodic situation, are analytically established in the asymptotic limit within an exact disorder matrix formalism developed using the uncorrelated binary nature of the coin-toss aperiodicity.

\end{abstract}
\maketitle

\section{Introduction}
Study of periodically driven closed quantum systems within the framework of the Floquet theory  is one of the most prominent areas of ongoing research in the context of non-equilibrium
dynamics (for review see, [\onlinecite{alessio13,eckardt15, bukov16}]). Floquet engineering \ct{oka18} has not only made it possible to prepare systems in hard to access equilibrium phases, but it can
also generate novel non-equilibrium phases of matter those have no equilibrium analogue. Recently discovered spontaneous time-
translational symmetry breaking time crystals\cite{else16, khemani16}, light-induced non-equilibrium superconducting and topological systems \cite{fausti11,rechtsman13},
Floquet graphene \ct{oka09,kitagawa10} and topological insulators \ct{lindner11}, dynamical generation of Floquet Majorana modes\cite{thakurathi13} are few such intriguing examples. 

On the other hand, the non-equilibrium
dynamics of periodically driven closed quantum systems also address the issues concerning the fundamental statistical aspects especially from the viewpoint of  thermalisation\cite{russomanno12,lazarides14}, dynamical freezing \ct{das10}, dynamical
localisation \ct{nag14,nag15,agarwala16}, etc.. 
Interestingly, under periodic drives the asymptotic behaviour of the system
depends crucially upon the integrability  or non-integrability of the underlying system. A periodically driven  system of   free fermions stops absorbing energy in the asymptotic limit of driving \ct{russomanno12}  and   reaches a periodic state which can be viewed as   a periodic generalized Gibbs ensemble with an extensive number of
stroboscopically conserved quantities \ct{lazarides14}. For a non-integrable system, on the other hand, the system is believed to absorb energy indefinitely
and reach the infinite temperature ensemble (ITE)  \ct{alessio14}. For a class of (nearly integrable) systems, however, there is a possibility of pre-thermalisation \ct{abanin17}.
Further,  a periodically driven non-integrable system does not necessarily thermalise rather may reach a 
 many-body localized phase \ct{ponte15}. It has also been reported that a many-body localised phase may get destabilised under a periodic driving \ct{lazarides15}.

Although the statistical nature of the
periodic steady states attained depending upon the integrability of the system in question, has become  of great interest to a
growing community,  recently, there have been several studies which probe the role of aperiodicity on the dynamics and the emergent steady states \ct{nandy17,bhattacharya17}. Exploring the consequences of breaking the time periodic structure of the drive with an aperiodic noise, for free
fermionic systems, it was shown that the system heats up to an ITE; however, for a self-similar disorder distribution, the resulting unitary dynamics leads to new emergent steady states, for instance, geometric
generalized Gibbs ensemble \ct{nandy17}. The time periodicity in periodic driving can also be broken using a biased coin-toss (or biased random walk) protocol: A Jordan-Wigner solvable
free fermion  system is again
found to heat up to infinite temperature in the asymptotic limit of driving even for the slightest deviation from perfect periodicity. Further, the asymptotic value of the residual
energy can be derived within an exact analytical disorder matrix ($D$-matrix) framework \ct{bhattacharya17}. Furthermore, this $D$-matrix formalism can determine the behavior of local and global observables clearly separating out the universal and non-universal features those coalesce to generate an intriguing interplay.  We also note in passing that it was established long back \ct{ott84,guarneri84}   that the expectation value of the kinetic
energy operator of a noisy $\delta$-perturbed single quantum rotator grows in time in an unbounded fashion, whereas in the present models we consider, the residual energy reaches a finite value in
the asymptotic limit.

%In the coin-toss protocol the breaking  of time periodicity  has been modeled to incorporate
%the effect of noise in the free fermionic system as a biased random walk (or a biased coin toss) scenario.
% But in contrast to earlier works, to study the dynamics of such a system at all times, it was necessary to formulate an exact framework . It not only addresses a fundamental question of thermalisation in an aperiodic closed quantum system from a statistical viewpoint but also quantitatively ascertains the host of non-universal behaviors those occur in the approach towards the infinite temperature ensemble (ITE).

Our goal in this work is not only to study the properties of Jordan Wigner integrable systems once it reaches an ITE but also to quantitatively study the approach of the system towards the ITE by looking at the behavior of certain local operators like the current and the residual energy in the presence of a coin-toss like aperiodic  driving. One-dimensional hard core Bosonic (HCB) systems, those capture the quantum phase transition from a superfluid (SF) to a Mott insulator (MI) phase and have been experimentally realized by trapping ultracold atoms in optical lattices \ct{paredes04, kinoshita04, klich07}, being analytically tractable, provide the perfect platform for studying the aperiodic dynamics of the above mentioned local quantities such as the current and the residual energy. Henceforth, one of  the main questions we address here is whether the initial current in the SF phase survive in the asymptotic limit $(t\to\infty)$ under these Floquet coin-toss kind of aperiodic perturbations even when the HCB chain is always in the SF phase except for the $\delta$-function kicks at aperiodic intervals.

The other fascinating question we address in this work is that of the entanglement entropy between different sub-parts of a large system because it measures quantum correlations in a more universal and straightforward way than do correlation functions captured in local observables such as currents and residual energies themselves.
The amount of entanglement in the ground state of a quantum many body system has been studied extensively in literature ~\ct{fazio8, latorre_r}. It is known that the entanglement entropy (EE) in the ground state of a short range $d$-dimensional quantum many body system ~\ct{kitaev3, calabrese04,latorre4, jin4, jin5} follows an area law $S_l \sim l^{d-1}$ upto a logarithmic correction ~\ct{wolf,gioev06}.
%where is $l$ is the subsystem size, i.e., $S_l$ depends only on the boundary of the subsystem. However, when the system is exactly at the critical point, the EE logarithmically diverges with the subsystem size as $S_l \sim \log (l^d)$. On the other hand,  away from the criticality, the scaling of the EE is dictated by  the finite correlation length $\xi$ ($\ll l)$ as $S_l \sim \log\xi^d $.

Interestingly, for a steady state reached by the system undergoing a non-equilibrium dynamics which is described by a finite temperature ensemble, the EE typically scales with volume of the subsystem \cite{calabrese5,calabrese8,sen16,russomanno16,apollaro16}, $S_l \sim l^d$ (with certain exceptions like in the case of many body localized systems \cite{berkovits12,bauer13}). For instance, following a sudden quench of a one-dimensional quantum many body system, $S_l$ initially increases linearly with time and then saturates to an asymptotic value which is proportional to the block size $l$ i.e. the volume of the  subsystem\cite{calabrese5,calabrese8}. The same kind of behavior in the EE is also observed for periodically driven one-dimensional  quantum many-body systems, but the asymptotic value that the EE attains is in this case ascertained from a periodic steady state described by a periodic generalized Gibbs ensemble \ct{sen16, russomanno16, apollaro16}. On the contrary, a disordered system or a state of a many body localized system has a characteristic logarithmic growth of entanglement in time \ct{calabrese6, burrell, prosen8, pollmann12, altman14, pollmann16}. Finally, following a quench of  a non-integrable  system with random initial  conditions, the average EE has been found to  increase linearly with time   and saturates to a thermal (infinite temperature) value of EE in the asymptotic limit of time \ct{huse13}.
 
 Let us summarise the main results of the paper at the outset:  in Sec. \ref{sec_cointoss} we elaborate on the Floquet coin-toss protocol and the $D$-matrix formalism deferring
 the calculational detail to the Appendix \ref{app_dmat}. In Sec. \ref{sec_HCB}, we introduce the model of hard-core bosons on a one-dimensional lattice in the presence of an aperiodically kicked (and also driven
 staggered potential)  and show that the system indeed reaches an 
  ITE in the asymptotic limit of driving where  the initial current vanishes for any frequency as soon as  a minimal aperiodicity is incorporated. The vanishing of the current is also illustrated 
  using an exact analytical approach based on the $D$-matrix formalism. In the process, we also compare with the corresponding perfectly periodic situation when for the $\delta$-kicking
  there is a dynamical localisation for which the current vanishes only in the large $\omega$ limit for asymptotic number of driving. Regarding the entanglement entropy ($S_l$) of the
  paradigmatic one dimensional transverse field  Ising model (TFIM) presented in Sec.~\ref{sec_ee}, we establish
  that, in contrast to the perfectly periodic situation,  the $S_l$ reaches the thermal value in the asymptotic limit for any frequency as a result of aperiodicity which can be analytically established using the $D$-matrix formalism. Remarkably, the initial ballistic growth of the $S_l$ and the maximum stroboscopic group velocity are found to be  robust against the aperiodic perturbation. In Appendix \ref{asymptotic_ee}, we 
  make recourse the D-matrix formalism and analytically establish that the $S_l$ indeed attains the thermal value $l$ for any driving frequency  in the asymptotic limit. Let us emphasise that although in subsequent sections,
  we shall present results choosing a particular value of $p$, the conclusions we draw are independent of choice of $p$ and are valid for any finite aperiodicity $p\neq 0, 1$.

%
%In this work, therefore, to study the dynamics of the entanglement entropy, we consider the paradigmatic one dimensional transverse field Ising model so that comparisons can easily be drawn between the present work and previous works those discuss the behavior of enetanglement entropy in this model in both non-equilibrium situations ~\cite{calabrese5,calabrese8,sen16,russomanno16,apollaro16}.

\section{Floquet coin toss dynamics}

\label{sec_cointoss}

To illustrate the essential idea behind the Floquet coin-toss problem, let us consider  a periodically driven two-level
system with Hamiltonian $H(t+T)=H(t)$,  initially prepared in the state $\ket{\psi(0)}$.  We revisit the disorder matrix formalism which provides an exact analytical
framework when temporal binary disorder is introduced on top of the periodic driving \ct{bhattacharya17}.

For a perfectly periodic situation,
 it is convenient to define a time evolution operator over a complete time period ($T$)  as $\cal{F}$$(T)=\cal{T} $$ \exp \left( -i \int_{0}^{T}H(t) dt \right)$.  The stroboscopic dynamics of the system is governed by an effective static Hamiltonian $H_F$, the so called Floquet Hamiltonian, defined as $\mathcal{F}(T)=\exp(-iH_F T)$.  After $N$ stroboscopic intervals of time (i.e.,
 after $N$ complete periods), the state of the system  is  given by 
 \begin{eqnarray}
\ket{\psi(NT)} &=& \left[\mathcal{F}(T)\right]^N \ket{\psi(0)} \\ \nonumber
 &=& r^+e^{-i\mu^+ NT}\ket{\phi^{+}}+r^-e^{-i\mu^- NT}\ket{\phi^{-}},
\end{eqnarray}
where $\ket{\phi^{\pm}}$ are the eigenstates of the Floquet Hamiltonian $(H_F)$ with corresponding eigenvalues $\mu^{\pm}$ (called Floquet quasi-energies) and the overlaps $r^{\pm}=\bra{\phi^{\pm}}\psi(0)\rangle $.

In the Floquet coin-toss situation,~\ct{bhattacharya17} a parameter ($\gamma$) of the system is driven in time in the following way:
\begin{equation}
	\gamma(t)=\sum_{n=1}^{N} g_n f(t),
	\label{eq_protocol}
\end{equation}
where $f(t)=f(t+T)$, is a periodic function of time and $N$ is the number of complete periods. Notably, the parameter  $g_n$ is a binary random variable which takes the value either $1$ with probability (bias) $p$ or $0$ with probability $(1-p)$, drawn randomly from a binomial distribution. Consequently, $g_n=0$ refers to the free evolution of the system for the time interval $(n-1)T$ to $nT$ i.e., there is a finite  probability ($1-p$) that the
periodic drive is missing  over any complete time period. The uncorrelated binary randomness in the parameter $g_n$ represents a coin-toss like  stochastically periodic driving. The case with $p=1$ represents 
the perfectly periodic situation while $p=0$ implies the free evolution of the system.

In the coin-toss like aperiodic situation, we  can  define a generic time evolution operator describing the evolution of the system from  $(n-1)T$ to $nT$ given by,
\begin{equation}
\bold{U}(g_n)=\begin{cases}
\mathcal{F}(T), & \text{if $g_n = 1$},\\
U^{0}(T), & \text{if $g_n = 0$},
\end{cases}
\label{eq_genu}
\end{equation}
where  $U^0(T)= \exp(-i H^0 T)$ is the time evolution operator for the free Hamiltonian $H^0$.  The time evolved state at a time $t=NT$  can then be written as time ordered product of the $N$  generic time evolution operators as follows:

\begin{equation}
\lvert \psi(NT)\rangle = \bold{U}(g_N) \bold{U}(g_{N-1}). . .  . . . .  \bold{U}(g_2)  \bold{U}(g_1) \lvert \psi(0) \rangle.
\label{eq_psint}
\end{equation}
 Therefore, the expectation value of some   operator $\hat{O}$, time-independent or  periodic with a time period $T$, at time $t=NT$ is given by,
 
\begin{eqnarray} 
\langle \hat{O}(NT) \rangle & = & \langle \psi(0) \rvert \bold{U}^{\dagger}(g_1) \bold{U}^{\dagger}(g_2) .......  \bold{U}^{\dagger}(g_N) \nonumber \\
 &\times& \hat{O} \bold{U}(g_N)........\bold{U}(g_2) \bold{U}(g_1) \lvert \psi(0) \rangle.
 \label{eq_expt}
 \end{eqnarray}
  The quantum average defined in Eq.~\eqref{eq_expt} also needs to be averaged over disorder configurations. To achieve that within an analytical framework, we use the eigenstates of the initial Hamiltonian $H^0$ as basis states and arrive at the expression (see Appendix \ref{app_dmat}) of the configuration averaged expectation value:
%Now to find the average of this quantity over several configurations of $N$ number of generic evolution operators, we will , unlike our earlier work, where we have used the Floquet modes as basis. This leads to a great simplification in the calculation of entanglement entropy in the asymptotic limit. However, all the  properties of the emergent disorder matrix remains the same in this basis.
%The final expression of the configuration averaged  $\langle\hat{O}(NT)\rangle$ is given by (for detailed calculation see \cite{}),
\begin{widetext}
\begin{equation}
	\overline{\langle\hat{O}(NT)\rangle} = \sum_{j_0,j_N,i_0,i_N}^{} \langle \psi(0)\rvert j_0 \rangle \langle j_N \rvert \hat{O} \lvert i_N \rangle \langle i_0 \rvert \psi(0) \rangle  \Big[ {D}^N \Big]_{j_0 j_N i_0 i_N};
	\label{eq_davg}
\end{equation}
\end{widetext}
here,  $j_i=1,2$, $i_i=1,2, \forall i = 0, N$, and the states $\lvert 1 \rangle = \lvert\psi^{0}_{g}\rangle$ and  $\lvert 2 \rangle = \lvert\psi^{0}_{e}\rangle$ refer to the ground and excited states of the initial Hamiltonian $H^{0}$.  In this basis, the elements of the ($4\times4$)  disorder matrix $D$ assume the form:

\begin{widetext}
	\begin{equation}
	D_{j_1,j_2,i_1,i_2} \equiv \left((1-p) e^{iT\left(E_{j_{1}}-E_{i_{1}}\right)} \delta_{j_1,j_2} \delta_{i_1,i_2} + p \langle j_2 \rvert \boldmath{\cal{F}}^{\dagger}(T) \lvert j_1 \rangle \langle i_1 \rvert \boldmath{\cal{F}}(T) \lvert i_2 \rangle \right),
	\label{eq_dmat}
	\end{equation} 
\end{widetext}
where $E_1$ ($E_2$) is the ground (excited) state energy of $H^0$. The above analytical approach which deals with the uncorrelated binary disorder in driving of such two level systems naturally leads to the emergence  of a  $4 \times 4$ disorder matrix $D$. Given the form  of  ${\mathcal F} (T)$ and the knowledge of disorder encoded in the bias $p$ of driving, every element of the $\boldmath{D}$-matrix can be calculated exactly. In the Appendix \ref{app_dmat}, we also established the structure of the $D$-matrix in the asymptotic limit of $N$ analysing the eigenvalues of the 
same,  which assumes the following simple form,
\begin{equation}
\lim_{N\to\infty}D^N=\frac{1}{2}\left(
\begin{array}{cccc}
1 & 0 & 0 & 1 \\
0 & 0 & 0 & 0 \\
0 & 0 & 0 & 0 \\
1 & 0 & 0 & 1 \\
\label{eq_dninf}
\end{array}
\right).
\end{equation}
Notably, this asymptotic structure of the $D$-matrix is universal and  independent of the driving  protocol ($\delta$-kicked or sinusoidal driving), its amplitude and
frequency  as well as the amount of the bias of coin-toss  $p$. As we elaborated in the Appendix \ref{app_dmat}, the $D$-matrix has four eigenvalues:  one  of them is unity, the other one is real having value less than unity while the  other two are complex conjugates of each other with magnitude less than unity. Although the eigenvalue unity dictates the asymptotic universal behaviour, the other eigenvalues having modulus less than unity vanish in the asymptotic limit while solely  determining all the non-universal early time behaviour of the quantities of interest. 

We emphasise that although for a general many-body system the indices  of the $D$-matrix can run over $2^N$ values, in this work we shall focus  on free fermionic many-body quantum systems those can be decomposed into decoupled two-level systems for each conserved quasi-momenta mode $k$ and hence the  indices run over two values only  throughout the rest of our work.

\section{Hard-core Bosonic system and current}

\label{sec_HCB}
 
\subsection{The 1D Hard Core Bosonic Chain}

\begin{figure}[]
\centering
\includegraphics[width=.42\textwidth,height=7cm]{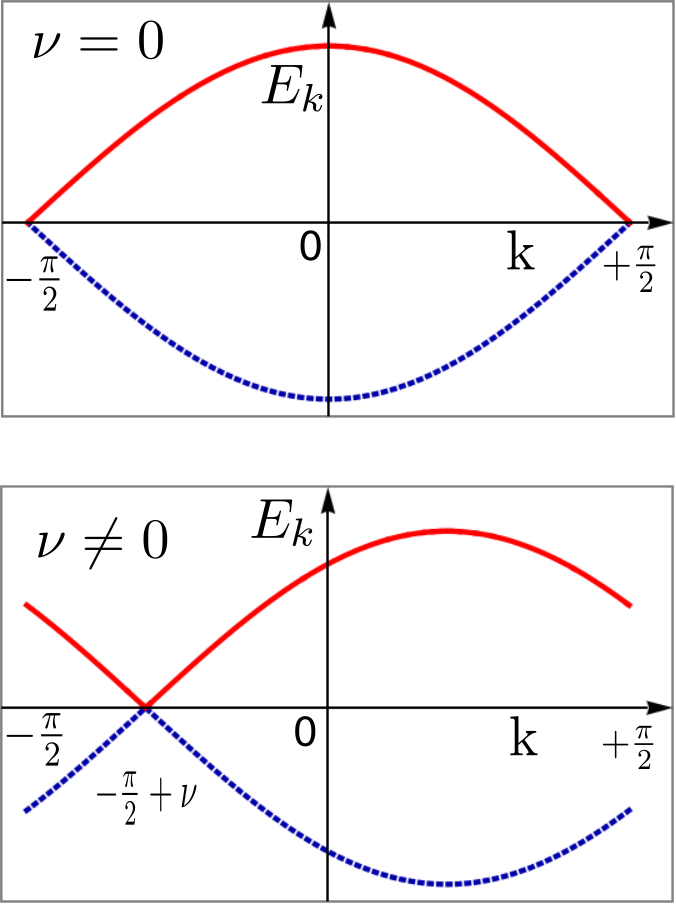}
		\caption{ (Color online) { Brillouin zone of the HCB chain with zero staggered potential showing the ground state (blue dashed line) state  and the excited (red solid line) state for the twist $\nu=0$ (top) and $\nu \neq 0$ (bottom).
		}}
		\label{fig_BZ}\end{figure}

In this section, we shall focus on  a one dimensional chain of hard core bosons (HCB) on a lattice of $L$ sites described by the Hamiltonian
\begin{equation}\label{tightb}
	H= -t \sum_{i=1}^{L}\left( b_i^{\dagger} b_{i+1} + b_{i+1}^{\dagger}b_i\right)
\end{equation}
where $t$ is the hopping amplitude (henceforth, scaled to unity) and $b_i$($b^{\dagger}_i$) are the bosonic annihilation (creation) operators satisfying the periodic boundary condition $b_{L+1}=b_1$ with the additional on-site hardcore conditions $(b_i)^2=(b_i^{\dagger})^2=0$ and $\{b_i,b_i^{\dagger}\} =1 $ which  forbids the  double occupancy. 
Applying Jordan-Wigner transformation $b_i = \prod_{l=1}^{i-1} \exp \left( i \pi c_l^\dagger c_l\right) c_i$ and $b_i^\dagger = c_i^\dagger\prod_{l=1}^{i-1} \exp \left( -i \pi c_l^\dagger c_l\right)$, where $c_i$ is the fermionic annihilation operator, the bosonic system described in Eq.~\eqref{tightb} can be mapped on to a noninteracting spin-less fermionic system given by the
Hamiltonian:
 \begin{equation}\label{tightbf}
 H= - \sum_{i=1}^{L}\left( c_i^{\dagger} c_{i+1} + c_{i+1}^{\dagger}c_i\right)
 \end{equation}
 %These fermions obey periodic or anti-periodic boundary condition depending on whether the number of sites are odd or even, as $.
 Considering an anti-periodic boundary condition on fermions $c_{L+1}=-c_1$,  the Hamiltonian gets decoupled in the Fourier space as $ H=-2t\sum_{k}^{} \cos k (c_k^{\dagger}c_k) $, where the allowed quasi-momenta are given by $k=2\pi m/ L$ with $m=-{(L-1)}/{2}, ..., -{1}/{2}, {1}/{2}, ..., {(L-1)}/{2}$. 
Introducing a pseudo spin basis $\ket{k}=c_k^\dagger \lvert0\rangle \equiv(1,0)^{T}$ and $\ket{k+\pi}=c_{k+\pi}^\dagger\lvert0\rangle\equiv(0,1)^{T}$, we can recast the Hamiltonian into a $2\times2$ form for each mode $k$ in the range $-\pi/2 \geq k \geq \pi/2$:
\begin{equation}
	H_k=-2\cos k \sigma_z,
\end{equation}
where $\sigma_z$ is the Pauli matrix. Clearly, the ground state of the system for each allowed $k$ mode is $(1,0)^{T}$.

When the system defined in Eq.~\eqref{tightb} is subjected to  an on-site staggered potential varying between $+V$ and $-V$ in alternate sites,  a coupling between $\ket{k}$ and $\ket{k+\pi}$ is generated. For each mode $k$, the form of the $2\times2$ Hamiltonian becomes,
\begin{equation}\label{stagh}
	H_k=-2\cos k \sigma_z + V\sigma_x.
\end{equation}
Consequently a gap opens up in the spectrum  at $k=\pm \pi/2$ for any nonzero value of $V$ and a phase transition occurs from a gapped Mott-insulator to a gap-less superfluid phase at $V=0$ \ct{klich07}.

Let us now consider  a boosted form of the Hamiltonian in Eq. \eqref{tightb} \ct{klich07,nag14,dutta15},
\begin{equation}\label{boostb}
	H_\nu= - \sum_{i=1}^{L} \left(b_i^{\dagger} b_{i+1}e^{-i \nu} + h.c.\right),
\end{equation}
which amounts to shifting momentum $k$ to $k-\nu$. The boosted Hamiltonian through gauge transformation can be made equivalent to a Hamiltonian with a twist in the boundary condition \ct{nag14}.
In this case,the ground state for individual $k$ mode is given by $(0,1)^T$ for $-\pi/2\geq k \geq -\pi/2+\nu$ and $(1,0)^T$ for $-\pi/2+\nu\geq k \geq -\pi/2$ as shown in Fig.~\ref{fig_BZ}. This asymmetry leads to  a finite non-zero amount of current in the ground state of the system. Introducing the current operator:
\begin{equation}
\hat{j}= -\frac{1}{L}\left(\frac{\partial H_\nu}{\partial\nu}\right)_{\nu=0}=(i/L)\sum_{i=1}^{L}(b_i^{\dagger} b_{i+1} - b_{i+1}^{\dagger}b_i),
\end{equation}
we find that for each $k$ mode the current operator assumes the form $\hat{j_k}=(2/L)\sin k \sigma_z$. Taking the average over the initial  ground  state as shown in Fig. ~\ref{fig_BZ} with $\nu \neq 0$,  the total initial current of the system in the thermodynamic limit ($L \rightarrow\infty$) is found to be  $j=(2/\pi)\sin \nu$.
\begin{figure*}[t]
	\centering
	\subfigure[]{%
		\includegraphics[width=.45\textwidth,height=5cm]{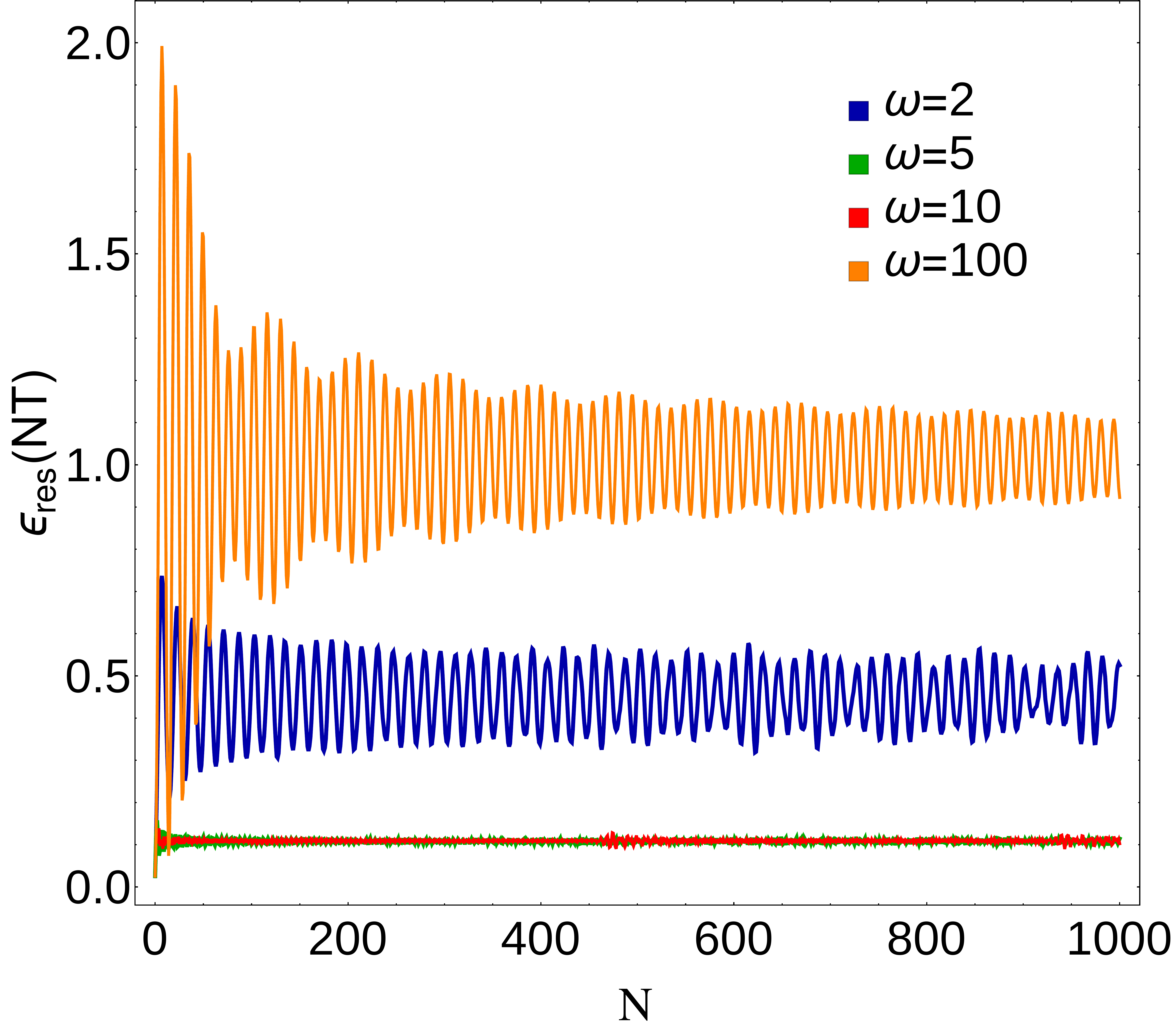}
		\label{fig_delta1}}
	\hspace{0.1cm}
	\quad
	\subfigure[]{%
		\includegraphics[width=.45\textwidth,height=5cm]{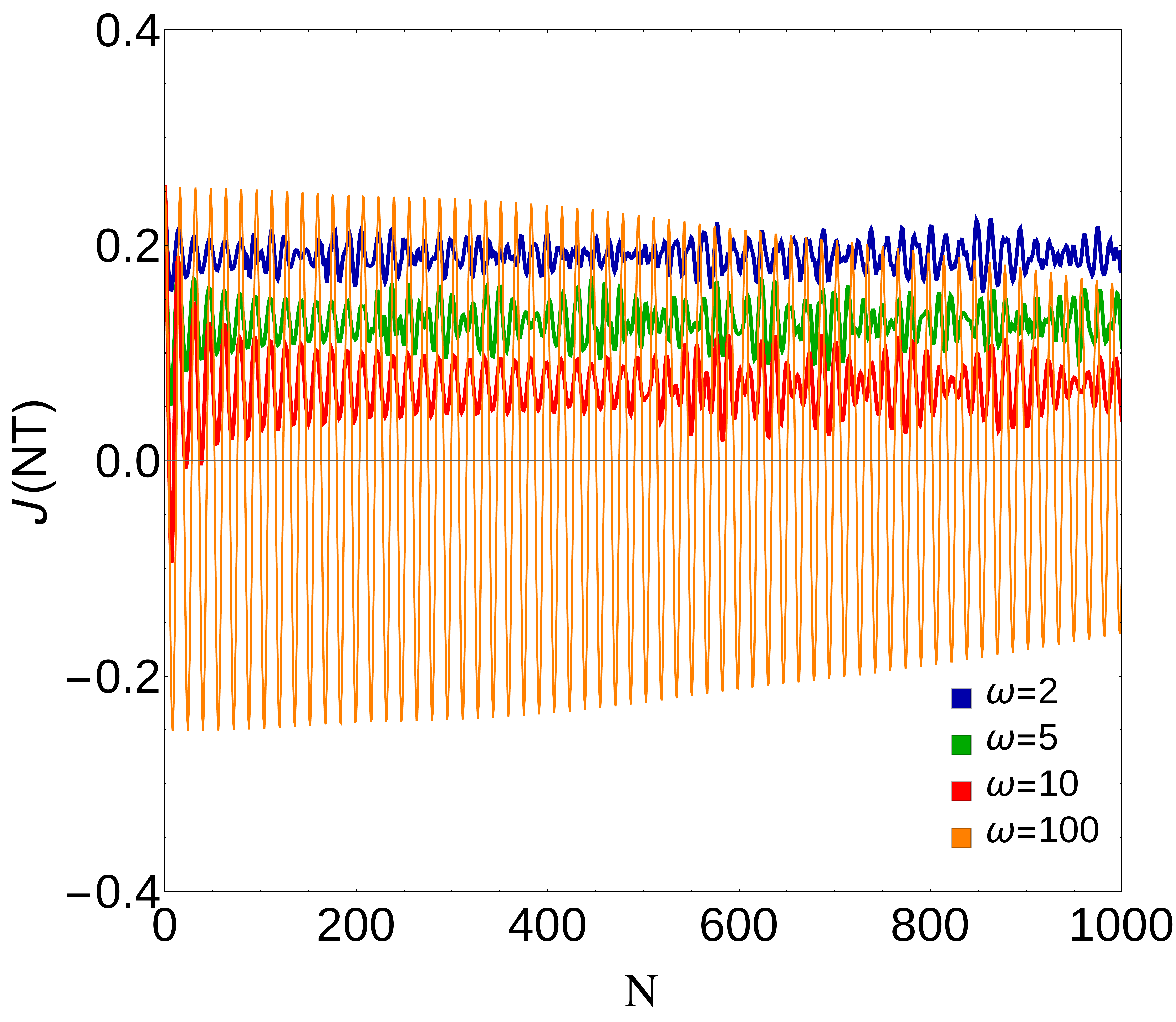}
		\label{fig_delta2}}
	\\
	\vfill
	\subfigure[]{%
		\includegraphics[width=.45\textwidth,height=5cm]{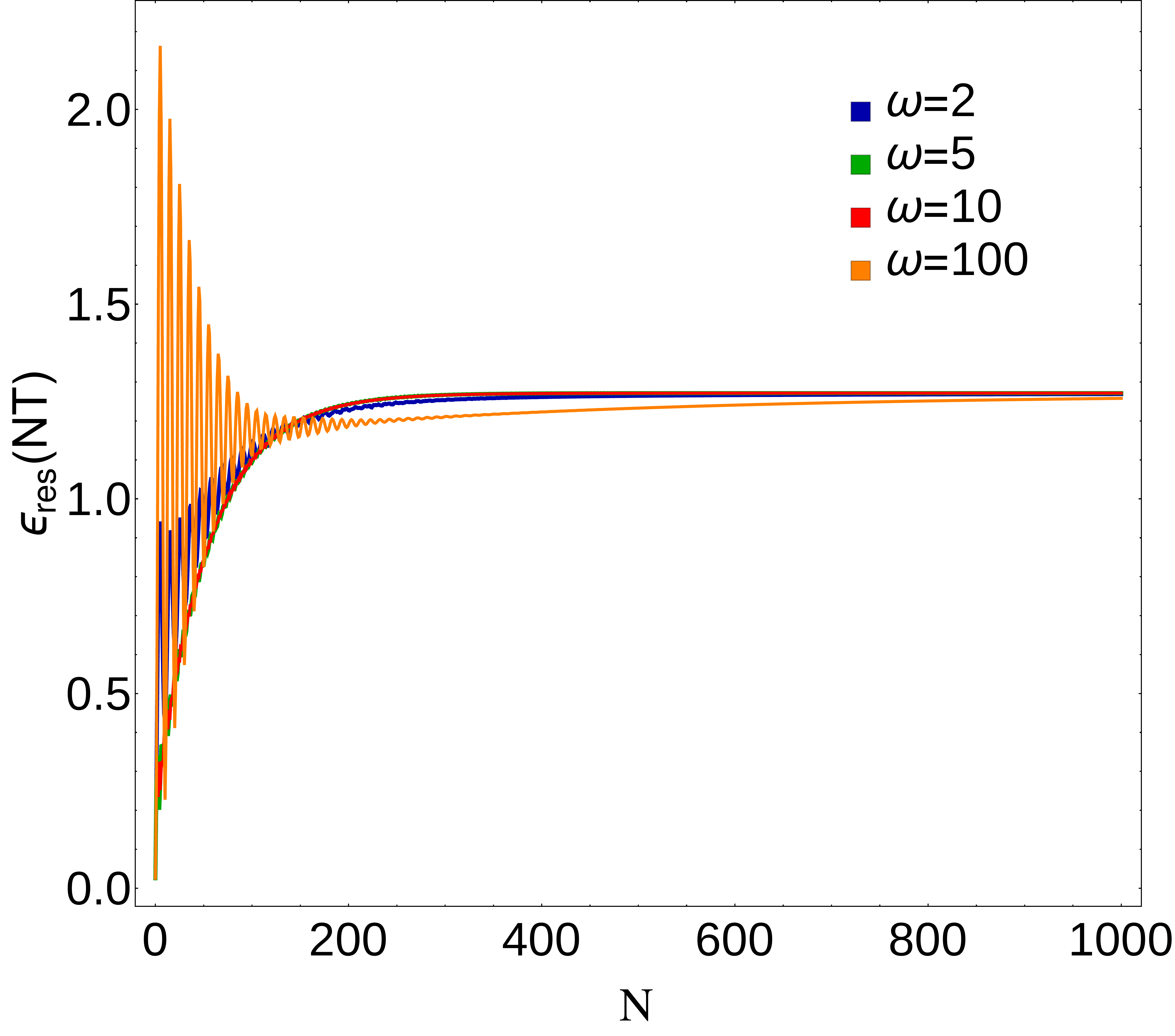}
		\label{fig_delta3}}
	\hspace{0.1cm}
	\quad
	\subfigure[]{%
		\includegraphics[width=.45\textwidth,height=5cm]{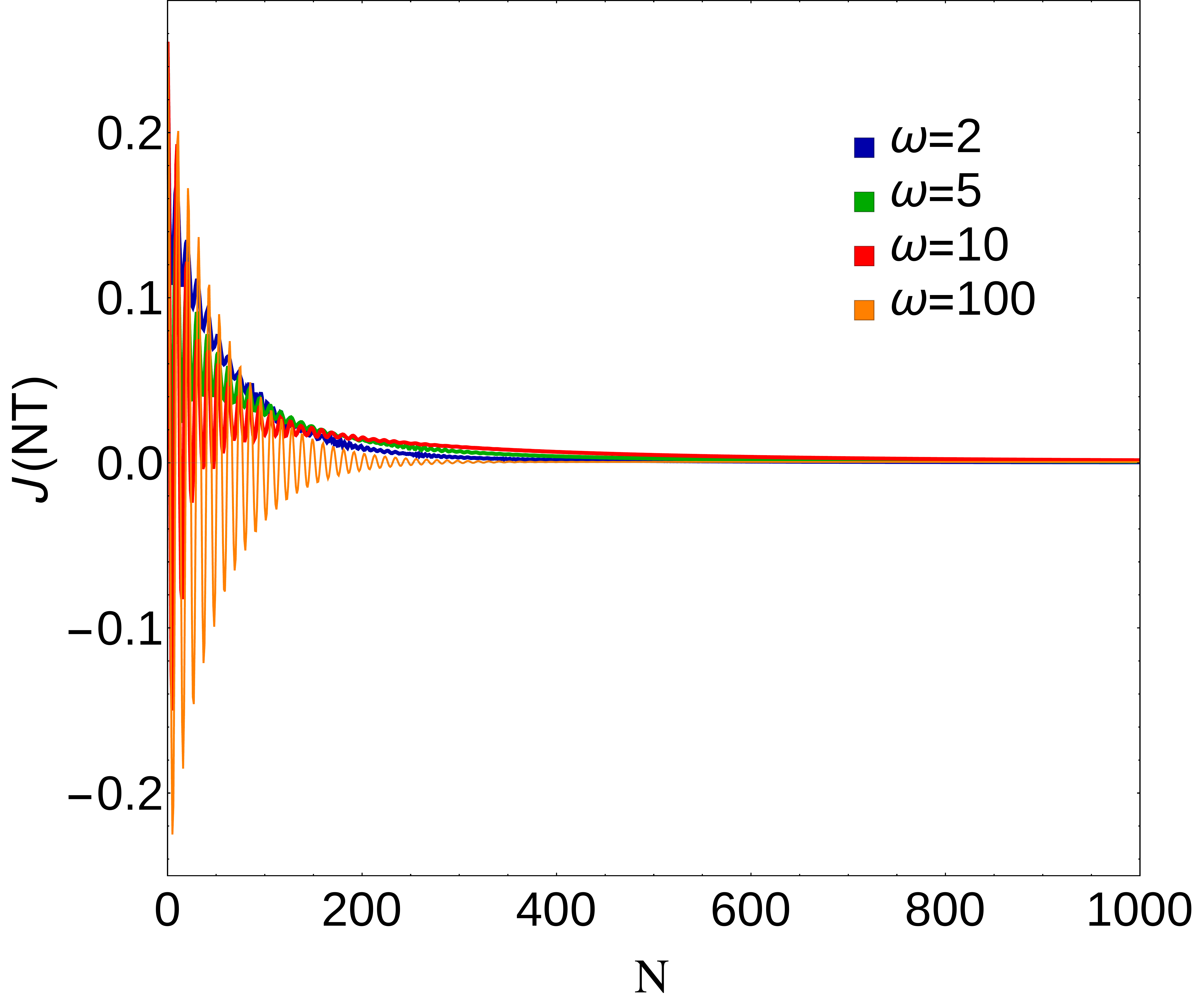}
		\label{fig_delta4}}
	\caption{ (Color online) {(a) The residual energy (RE) and (b) current ($ J(N)$) plotted as a function of stroboscopic intervals ($N$) for HCB chain with periodically kicked ($p=1$)  staggered potential with several values of frequency ($\omega$). Both the RE as well as the current reach  frequency dependent steady state values. The mean value of the current decreases monotonically with increasing frequency and  vanishes in the asymptotic limit of $N$ for very large value of $\omega$ resulting in the {\it dynamical localisation}. In the Fig.~(c) and Fig.~(d), we plotted the disorder averaged RE and current, respectively, with aperiodically kicked (with bias $p=0.5$) staggered potential. We observed that for all the frequencies RE goes to a universal value (corresponding to  the infinite temperature) while the current vanishes in the asymptotic limit of $N$. Here, the amplitude of kicking $\alpha=\pi/16$, twist parameter $\nu = 0.2$, and the system size $L=1000$ and we observe similar behaviour for all values of $p \neq 0,1$. In the case of aperiodic drive, the current vanishes very fast as a function of $N$ compared   to the dynamical localisation situation in the periodic high frequency $\delta$-kicking.
		}}
		\label{fig_delta}
	\end{figure*}

\subsection{The Current and The Residual Energy under periodic and aperiodic driving}

\label{sec_delta}

 After preparing the initial state of the system as the current carrying  ground state of the twisted Hamiltonian, we subsequently remove the twist (i.e., set $\nu=0$) at $t=0$. We then explore the behaviour  of the residual energy of the system and the initial current under the application of a perfectly periodic drive and coin-toss like aperiodic drive of the staggered potential choosing  the protocol given in Eq.~\eqref{eq_protocol} with  $\gamma(t)=V(t)$. 
 Let us first consider the {\it perfectly periodic} situation with $p=1$, considering two driving protocols with amplitude $\alpha$:  (i) delta kicking of the staggered potential , $ f(t) = \sum_{n=1}^{N} \alpha \delta(t-nT)$ and (2) sinusoidal driving, $f(t)=\alpha\sin(2\pi t/T)$, where $\alpha$ is the amplitude of driving. Let us note here that for the periodic delta kicking, we can find an analytical form of the Floquet evolution operator, 
 \begin{equation}
 \mathcal{F}_k(T)=e^{-i\alpha\sigma_x}e^{i2T\cos k \sigma_z}, 
 \end{equation}
 whereas, for sinusoidal driving  the same can only be obtained numerically. 	
 
 \begin{figure*}[]
		\centering
		\subfigure[]{%
			\includegraphics[width=.45\textwidth,height=5.0cm]{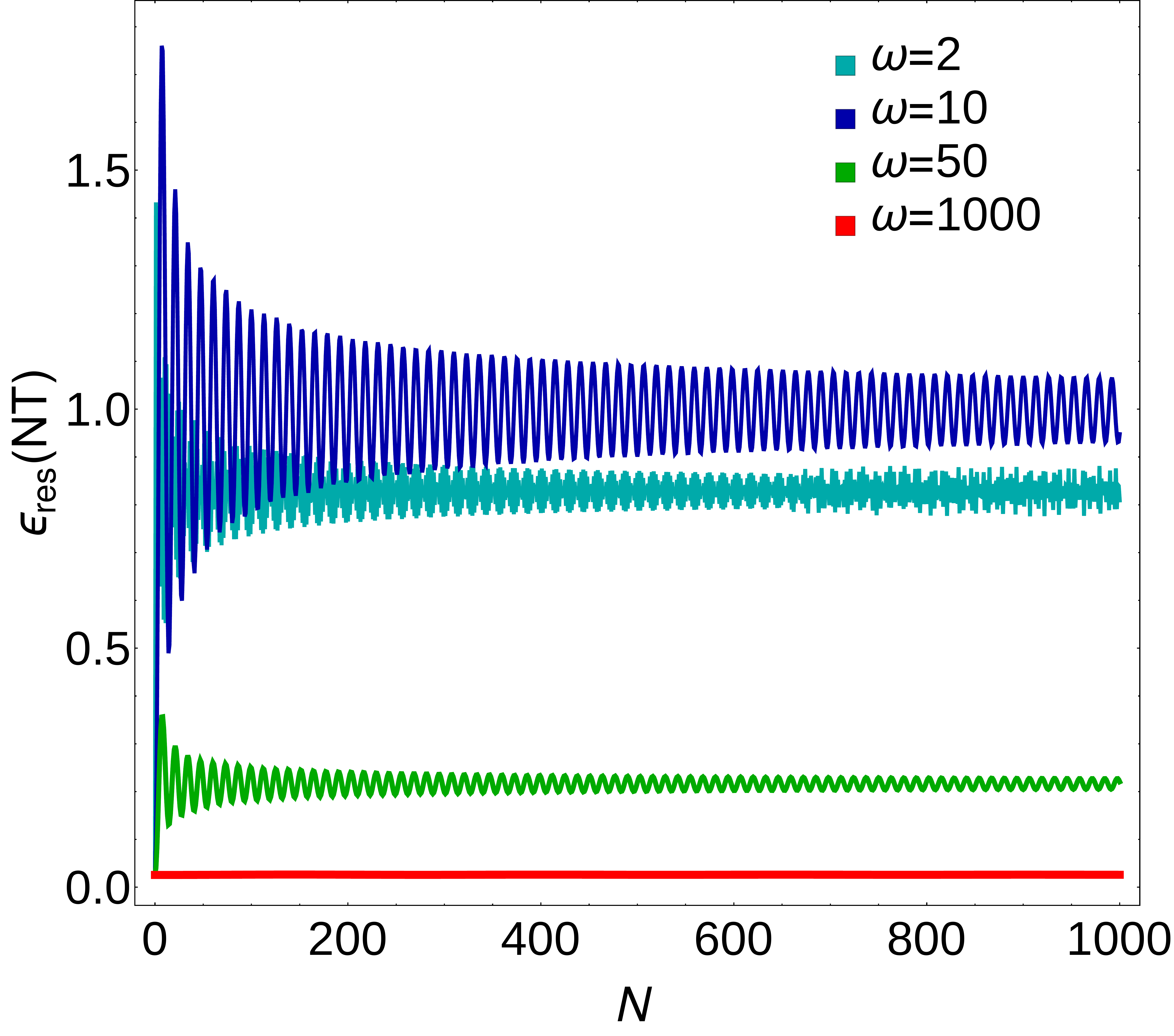}
			\label{fig_sin1}}
		\quad
		\subfigure[]{%
			\includegraphics[width=.45\textwidth,height=5.0cm]{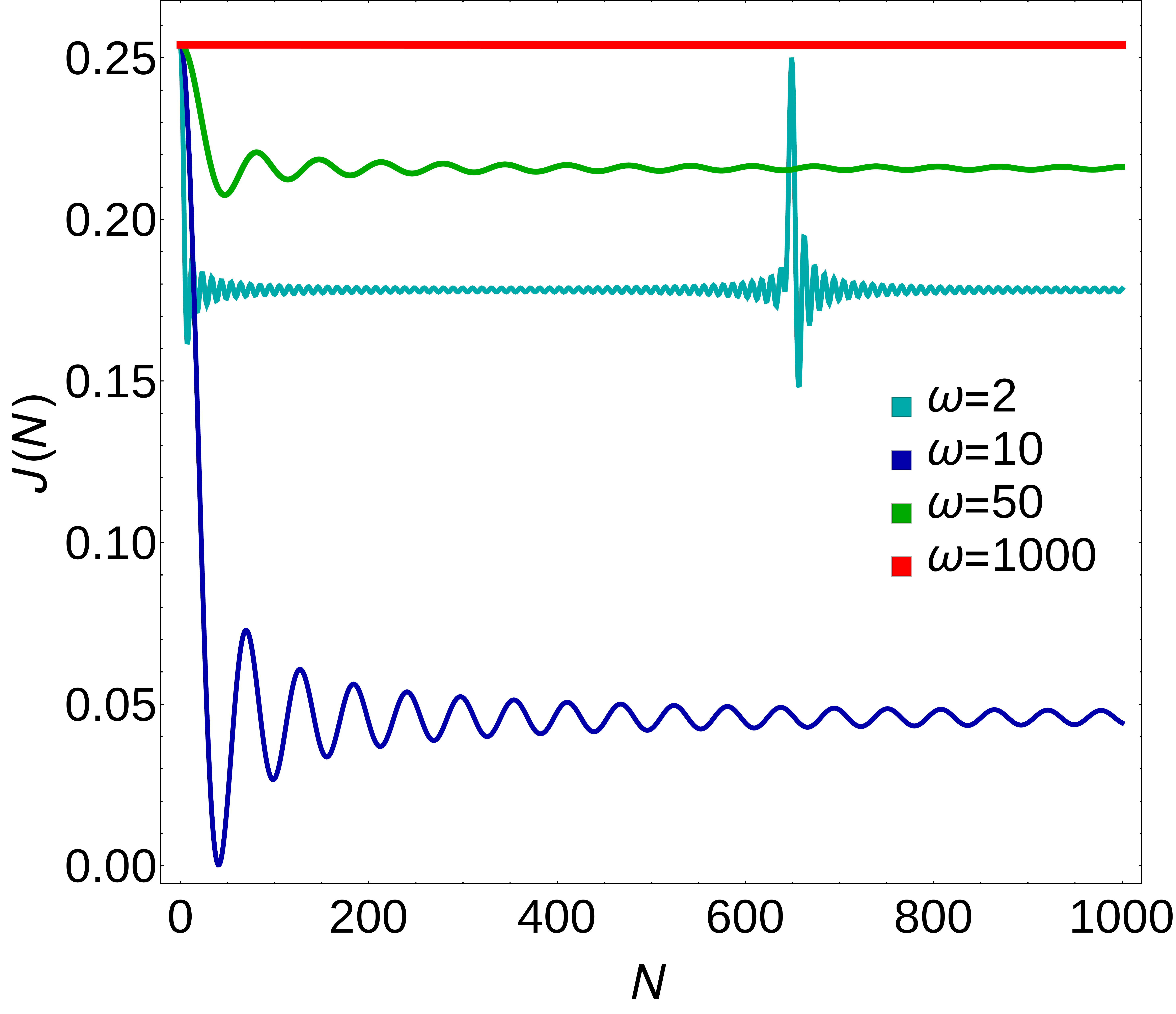}
			\label{fig_sin2}}
		\\
		\vfill
		\subfigure[]{%
			\includegraphics[width=.45\textwidth,height=5.0cm]{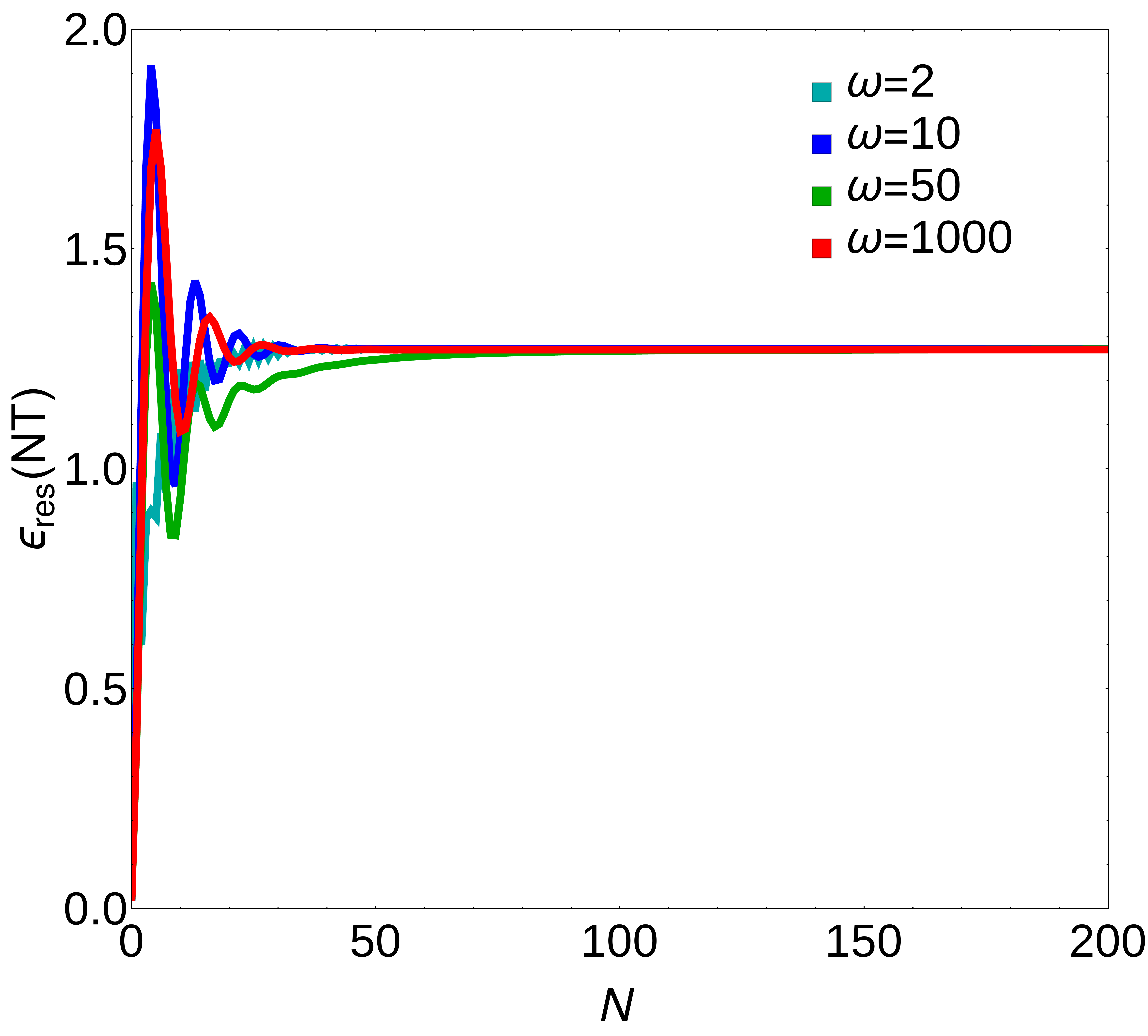}
			\label{fig_sin3}}
		\quad
		\subfigure[]{%
			\includegraphics[width=.45\textwidth,height=5.0cm]{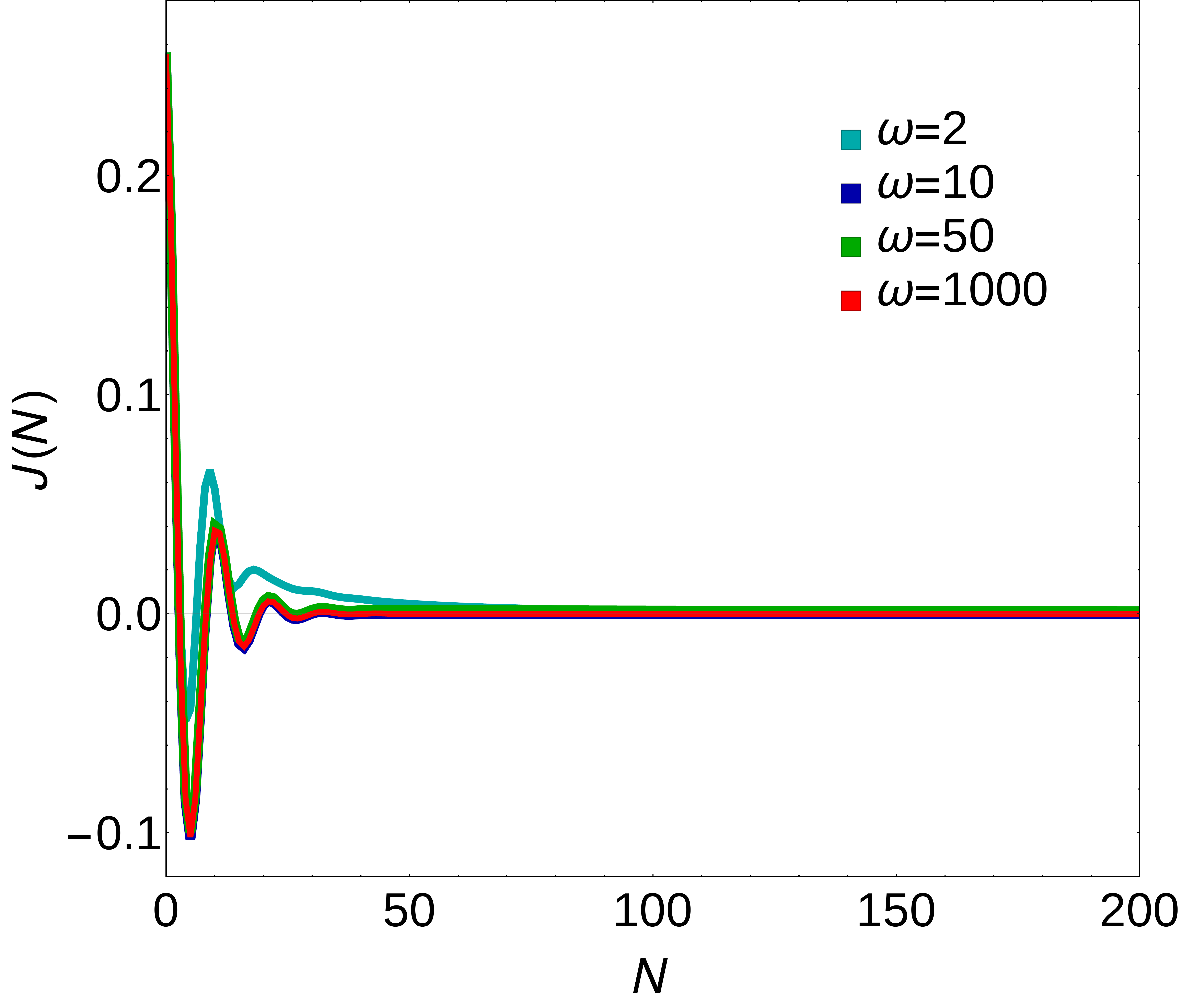}
			\label{fig_sin4}}
		\caption{ (Color online) {(a) The residual energy (RE) and (b) the current ($ J(N)$) plotted as a function of stroboscopic intervals ($N$) for HCB chain with perfectly periodic ($p=1$) sinusoidal driving of the staggered potential.  Unlike the delta kicking, both the RE and the current stick to their initial value for very high frequency of driving resulting implying the absence of dynamical localization. In the Fig.~(c) and Fig.~(d), we plotted the disorder averaged RE and current, respectively, for aperiodically sinusoidal driving (with bias $p=0.5$) of staggered potential. We also observed the saturation of RE to infinite temperature value and vanishing of the current in this case for all values of $\omega$ similar to the aperiodic delta kicked situation. Here, we have chosen amplitude of driving $\alpha=5$, twist parameter $\nu = 0.2$, and the system size $L=1000$. 
			}}
			\label{fig_sin}
		\end{figure*}

The residual energy (excess energy over the ground state) after $N$ complete period of kicking/driving  is given by $\epsilon_{res}(NT)=\frac{1}{L} \sum_k \left( e_k(NT) - e_k(0) \right)$, where $e_k(NT)=\langle\psi_k(NT) \vert H_k^0 \rvert \psi_k(NT)\rangle$  and $e_k(0) = \langle\psi_k(0) \vert H_k^0 \rvert \psi_k(0)\rangle$; here $H_k^0$ is the initial Hamiltonian.   The expression of current after $N$ complete periods is   $J(NT)=\sum_k \langle \psi_k(NT) \lvert\hat{j_k}\rvert \psi_k(NT)\rangle $. Therefore, evaluation of both the residual energy and current  involves calculation of  $\langle \psi_k(NT) \lvert\sigma_z\rvert \psi_k(NT)\rangle$ for individual $k$ mode. 

For perfectly periodic case ($p=1$), 

\begin{align} \label{psigma_avg}
&\langle \psi_k(NT) \lvert\sigma_z\rvert \psi_k(NT)\rangle= &&\\ 
& \hspace{1.0cm}\sum_{m, n =\pm} r_k^{m *}r_k^n e^{i \left( \ \mu_k^m - \mu_k^n  \right) NT} \langle \phi_k^m\lvert \sigma_z\rvert \phi_k^m \rangle. \nonumber
 \end{align}
In the asymptotic limit, a  periodic steady state value of the residual energy and the current is given by the expression: 
	\begin{equation}
		\sum_{m=\pm} \lvert r_k^m\rvert^2 \langle \phi_k^m\lvert \sigma_z\rvert \phi_k^m \rangle,
		\label{psigma_dia}
	\end{equation} 
Here we have exploited the fact that  all the rapidly oscillating terms (with $m \neq n$ in Eq.~\eqref{psigma_avg}) decay to zero in this asymptotic limit of $N$ when summed over all the $k$ modes.

For perfectly periodic delta kicking, there is a saturation in the stroboscopic residual energy and vanishing of the initial current
in the asymptotic limit of driving ($N \to \infty$) in the limit of large $\omega$ (compared to the maximum band width). This means that the system reaches a periodic steady
state where it stops absorbing energy and the vanishing of the initial current can be attributed to the phenomena of  dynamical localization  as reported  already in Ref. \ct{nag14}
Diagonalising the Floquet evolution operator to obtain $|\phi_k^{\pm}\rangle$ and $r_k^{\pm}$  and using the  Eq.~\eqref{psigma_avg}, we evaluated the RE and current as shown in Fig.~\ref{fig_delta1} and Fig.~\ref{fig_delta2} choosing different values of the frequency.  Clearly, the system indeed reaches a periodic steady state when the RE approaches a frequency dependent constant value
as $N \to \infty$. Similarly,  current goes to zero for very high frequency (say, $\omega=100$)  as the transient oscillations  die out in the asymptotic limit; on the contrary, for the small values of frequency the current saturates to some  non zero frequency dependent value.
 
 Let us digress to the {\it aperiodic} situation with $p \neq 1$. In the Fig.~\ref{fig_delta3} and Fig.~\ref{fig_delta4}, we show the disorder averaged stroboscopic residual energy and current for the aperiodic kicking ($p\neq1$) for different frequency of the stroboscopic intervals. It is
 evident that for all the frequencies the RE saturates to a universal value and the current goes to zero in the asymptotic limit of $N$. 
These results  can be analytically shown using the disorder matrix formalism by calculating the averaged value of $\langle \psi_k(NT) \lvert\sigma_z\rvert \psi_k(NT)\rangle$ over several configurations. For this purpose, we use the Eq.~\eqref{eq_davg}, with operator $\hat{O} \equiv \sigma_z$. For each $k$ mode,  we then find

	\begin{eqnarray}\label{eq_dsigma}
			&&\overline{ \langle\sigma_z(NT)\rangle}_k  \equiv\overline{\langle \psi_k(NT) \lvert\sigma_z\rvert \psi_k(NT)\rangle} \\ 
			&&=\begin{cases}
				\sum_{i=1}^{2}(-1)^i\left[D_k^N\right]_{1,i,1,i}, & \text{for $-\frac{\pi}{2} \leq k \leq -\frac{\pi}{2}+\nu $},\\ \nonumber
				\sum_{i=1}^{2}(-1)^i\left[D_k^N\right]_{2,i,2,i}, & \text{for $-\frac{\pi}{2}+\nu \leq k \leq +\frac{\pi}{2} $},
			\end{cases}
	\end{eqnarray}
		where in the above two equations, the $1^{st}$ and $3^{rd}$ index of the $D^N$ are fixed to $1$ (in the first equation) and $2$ (in the second equation) by the initial state $\lvert\psi_k(0)\rangle=\lvert 1\rangle=(0,1)^T$ for $-\frac{\pi}{2} \leq k \leq -\frac{\pi}{2}+\nu $ (Region I) and  $\lvert\psi_k(0)\rangle=\lvert 2\rangle=(1,0)^T$ for $-\frac{\pi}{2}+\nu \leq k \leq +\frac{\pi}{2} $ (Region II) respectively. While, the other index of the remaining two indices is absent due to the relation $\langle i\lvert \sigma_z \rvert j \rangle=(-1)^i\delta_{ij}$. Now using Eq.~\eqref{eq_dsigma}, we can calculate the configuration averaged value of residual energy and current at time $t=NT$, which are given by,
		\begin{eqnarray}
			\overline{\epsilon_{res}(NT)} &=&  \frac{1}{L} \sum_{k} \left\{ - 2\cos k
			\overline{ \langle\sigma_z(NT)\rangle}_k -e_k(0)\right\}, \\
			\overline{J(NT)} &=&  \frac{1}{L} \sum_{k} \left\{  2\sin k
			\overline{ \langle\sigma_z(NT)\rangle}_k\right\}.
		\end{eqnarray}

In the asymptotic limit ($N\rightarrow\infty$), from the structure of the $N^{th}$ power of the $D$ matrix in the Eq.~\eqref{eq_dninf}, the quantity $\overline{ \langle\sigma_z(NT)\rangle}_k$  vanishes for each $k$ mode in both the Regions I and II. So in the asymptotic limit, the initial current in the system goes to zero. On the contrary, the residual energy of the system saturates to the value of residual energy of an infinite temperature ensemble and is given by $\overline{\epsilon_{res}(\infty)}=-1/L\sum_k e_k(0)= 1.272$.  Comparing the Fig.~\ref{fig_delta2} and Fig.~\ref{fig_delta4}, we observed that in the case of aperiodic $\delta$-kicking, the current vanishes very fast as function of $N$ compared to the dynamically localised situation in the periodic high frequency $\delta$-kicking. This is because in the aperiodic case, two complex conjugate eigenvalues of the $D$-matrix, responsible for the transient oscillations, have modulus less than unity and vanish faster in the large $N$ limit.

Let us now consider the case of sinusoidal driving. In the perfectly periodic case, i.e., $p=1$, we numerically diagonalise the Floquet operator and calculate the 
RE and the current using Eq.~\eqref{psigma_avg}: results are presented in Figures~\ref{fig_sin1} and \ref{fig_sin2}.  It is noteworthy that here in the high frequency limit  the system fails to respond to the drive at all, and both the RE and current stick to their initial values. This is because of the fact that in the high frequency limit the effective Floquet Hamiltonian  can be approximated as $H_k^F \sim (1/T)\int_{0}^{T} dt H_k(t)$ i.e., equal to the bare Hamiltonian $H_k^0$ as the integral over a complete period of $\sin (2\pi t/T)$ vanishes and therefore the effect periodic driving essentially disappears. This is in contrast to the case for  $p=1$ delta kick situation, where the time-averaged Hamiltonian has a non zero mean over a complete period of time and  there is a possibility of dynamical localisation in the large $\omega$ limit which is absent in the case of sinusoidal driving. However, in the aperiodic case $p\neq 1$  (Figures \ref{fig_sin3} and \ref{fig_sin4}), we establish that  the asymptotic universal nature of the stroboscopic RE and current is identical to the aperiodically $\delta$-kicked situation. Although numerical results are presented for $p=0.5$, this universal nature holds for any non-zero value of $p$. Our results therefore establish that the current vanishes in the asymptotic limit of the aperiodic driving for any frequency for
infinitesimal aperiodicity (i.e, for any $p \neq 1$). This vanishing of the current  is also expected from the $D$-matrix analysis discussed above in the context of aperiodic $\delta$-kicks as the asymptotic structure of
the $D$-matrix is independent of the driving protocol as discussed in Sec.~\ref{sec_cointoss}.

\section{Entanglement Entropy}
\label{sec_ee}

In this section, we shall calculate the entanglement entropy ($S_l$)   by dividing the one dimensional many-body system into two parts, a subsystem ($A$ with block size of linear dimension $l$) and the rest of the system ($B$ with block size $L-l$, where $L$ is the length of the total system) and computing the reduced density matrix ($\rho_l$) of the subsystem $A$. The block entanglement entropy $(S_l)$ is defined as,
\begin{equation}
	S_l = -{\rm Tr}\left[ \rho_l \log_2(\rho_l) \right].
	\label{eq_vnee}
\end{equation}

\subsection{The Transverse Field Ising Model (TFIM)}
For studying  the entanglement entropy, we shall use the paradigmatic one-dimensional TFIM  described by the Hamiltonian,   
\begin{equation}\label{eq:1}
H=-\frac{1}{2}\sum_{n=1}^{L}\left[\tau_n^{x}\tau_{n+1}^{x}+h(t)\tau_{n}^{z}\right]
\end{equation}
where $h(t)$ is the time dependent transverse field and $ \tau_{n}^{i}$ $\{i=x,y,z\}$ are the Pauli spin matrices defined at the $n^{th}$ site. This model can be  exactly solved  via a  Jordan-Wigner  mapping of spin
variables to spinless fermions $\left(c_n, c^{\dagger}_n\right)$ \ct{dutta15};
through the transformation relations, $c_n=\exp \left(\pi i \sum_{j=1}^{n-1}a_j^{\dagger}a_j\right)a_n$ and $c_n^{\dagger}=a_n^{\dagger}\exp \left(-\pi i \sum_{j=1}^{n-1}a_j^{\dagger}a_j\right)$ where $a_n = \frac{1}{2}\left(\tau_{n}^{x} - i\tau_{n}^{y}\right)$.  Using the translational invariance of the lattice, one employs a Fourier transformation,  $c_n=\frac{1}{\sqrt{L}}\sum_k c_k e^{ink}$ and  the Hamiltonian can be decoupled in to $2\times 2$ blocks for each Fourier mode  such that  $H=\sum_k H_k$ with,
$$ H_k =  (h(t)-\cos k)c_k^{\dagger}c_k + \frac{\sin k}{2} (c_k^{\dagger}c_{-k}^{\dagger}+c_kc_{-k}), $$
 where the allowed values of $k$ modes are $k=\frac{2m\pi}{L}$ with $m=-\frac{L-1}{2}, ..., -\frac{1}{2}, \frac{1}{2}, ..., \frac{L-1}{2}$ corresponds to the anti-periodic boundary condition for the fermions with even number of $L$. We note that, $H_k$ connects only the definite parity sector i.e. the vacuum of Jordan-Wigner fermions $\lvert 0 \rangle$ with $c_k^{\dagger} c_{-k}^{\dagger}\lvert0\rangle$ or $c_k^{\dagger}\lvert 0 \rangle$ with  $c_{-k}^{\dagger}\lvert 0 \rangle$. Furthermore considering the even parity sector, in the basis of $\lvert 0 \rangle$ and $c_k^{\dagger} c_{-k}^{\dagger}\lvert0\rangle$, $H_k$ can be written in terms of Pauli matrices as $H_k=(h(t)-\mathrm{cos}k)\sigma_z+\mathrm{sin}k\sigma_x $.

We recall that it has already been established that under an aperiodic driving (or kick) within a coin-toss protocol, the TFIM reaches the ITE; the residual energy has a bounded growth to an asymptotic value
\ct{bhattacharya17} similar to the HCB chain discussed in  Sec.~{\ref{sec_delta}. In the subsequent discussion, we shall probe the manifestation of ITE  in the asymptotic behaviour of entanglement entropy in the process exploring the nature of its initial growth in time. 

At $t=0$, we assume that the system is in the ground state of $H$ with $h(t=0)=h^0$. This ground state has the following BCS like form,
\begin{equation}
	\lvert\psi(0) \rangle = \prod_{k>0}^{} \left( v_k^0 + u_k^0c_k^{\dagger}c_{-k}^{\dagger}\right)\lvert0\rangle
\end{equation} 
with $v_k^0=\cos (\theta_k/2) $ and $u_k^0=-\sin (\theta_k/2) $ where the angle $\theta_k$ is given by $\tan \theta_k = (\sin k)/(h^0 -\cos k)$. As the dynamics does not mix the parity subspaces, the time evolved state can be written as,
\begin{equation}
\lvert\psi(t) \rangle = \prod_{k>0}^{} \left( v_k(t) + u_k(t)c_k^{\dagger}c_{-k}^{\dagger}\right)\lvert0\rangle
\label{eq_bcst}
\end{equation} 
where the coefficients $u_k(t)$ and $v_k(t)$ are the solutions of the Bogoliubov-de Gennes equation,
\begin{equation}
	i \hbar \frac{d}{dt} \left({\begin{array}{cc} u_k(t) \\ v_k(t) \end{array}}\right) = H_k(t)\left({\begin{array}{cc} u_k(t) \\ v_k(t) \end{array}}\right)
	\label{eq_bdg}
\end{equation}
with the initial condition $u_k(0)=u_k^0 $ and $v_k(0)=v_k^0$.

\subsection{Evolution of EE}

 To compute the time evolution of the EE of the same model, we introduce two $l\times l$ correlation matrices \ct{eisert08} $C$ and $F$ of the JW-fermions with the matrix elements $C_{mn}=\langle\psi(t)\rvert c_m^{\dagger} c_n\lvert \psi(t)\rangle$ and $F_{mn}=\langle\psi(t)\rvert c_m^{\dagger} c_n^{\dagger}\lvert \psi(t)\rangle$ respectively, where $ 1 \leq m,n\leq l $. Finally, EE can be calculated from the Von Neumann entropy of a $2l \times 2l$ correlation matrix defined as,
\begin{equation}
	\bold{\cal{C}}_l (t) = \left({\begin{array}{cc} I-C & F \\ F^{\dagger} & C \end{array}}\right)
	\label{eq_cormat}
\end{equation}
where $I$ is the $l\times l$ identity matrix. Now the entanglement entropy of the sub-block of size $l$ is given by,
\begin{equation}
	S_l(t)=-{\rm Tr}\left[\bold{\cal{C}}_l (t) \log \bold{\cal{C}}_l (t)\right]=-\sum_{i}^{2l} \lambda_i(t) \log \left[\lambda_i(t)\right]
	\label{eq_vnecormat}
\end{equation}
where $\lambda_i(t)$ are the eigenvalues of the correlation matrix $\bold{\cal{C}}_l (t)$, and they come in pairs of $\lambda(t)$ and $1-\lambda(t)$. We note that
the  EE computed from the correlation matrix is  exactly equal to that obtained from the  Von Neumann entropy of the reduced density matrix $\rho_l$ given in Eq.~\eqref{eq_vnee}. 

\subsection{Results}

\begin{figure*}[]
		\centering
		\subfigure[]{
		\includegraphics[width=0.43\textwidth,height=5cm]{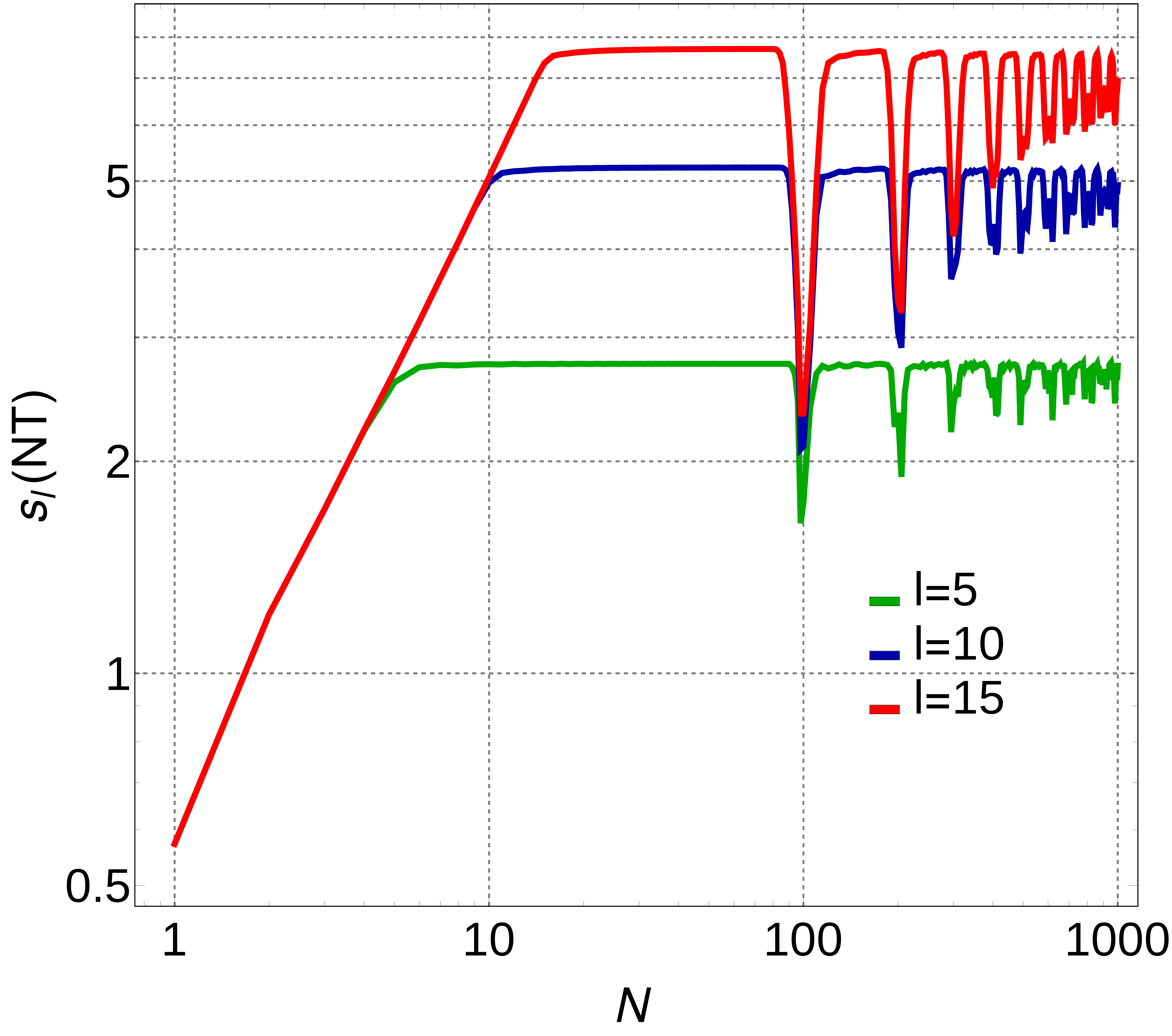}
		\label{fig_peevn}}
		\quad
		\subfigure[]{%
			\includegraphics[width=.42\textwidth,height=5.0cm]{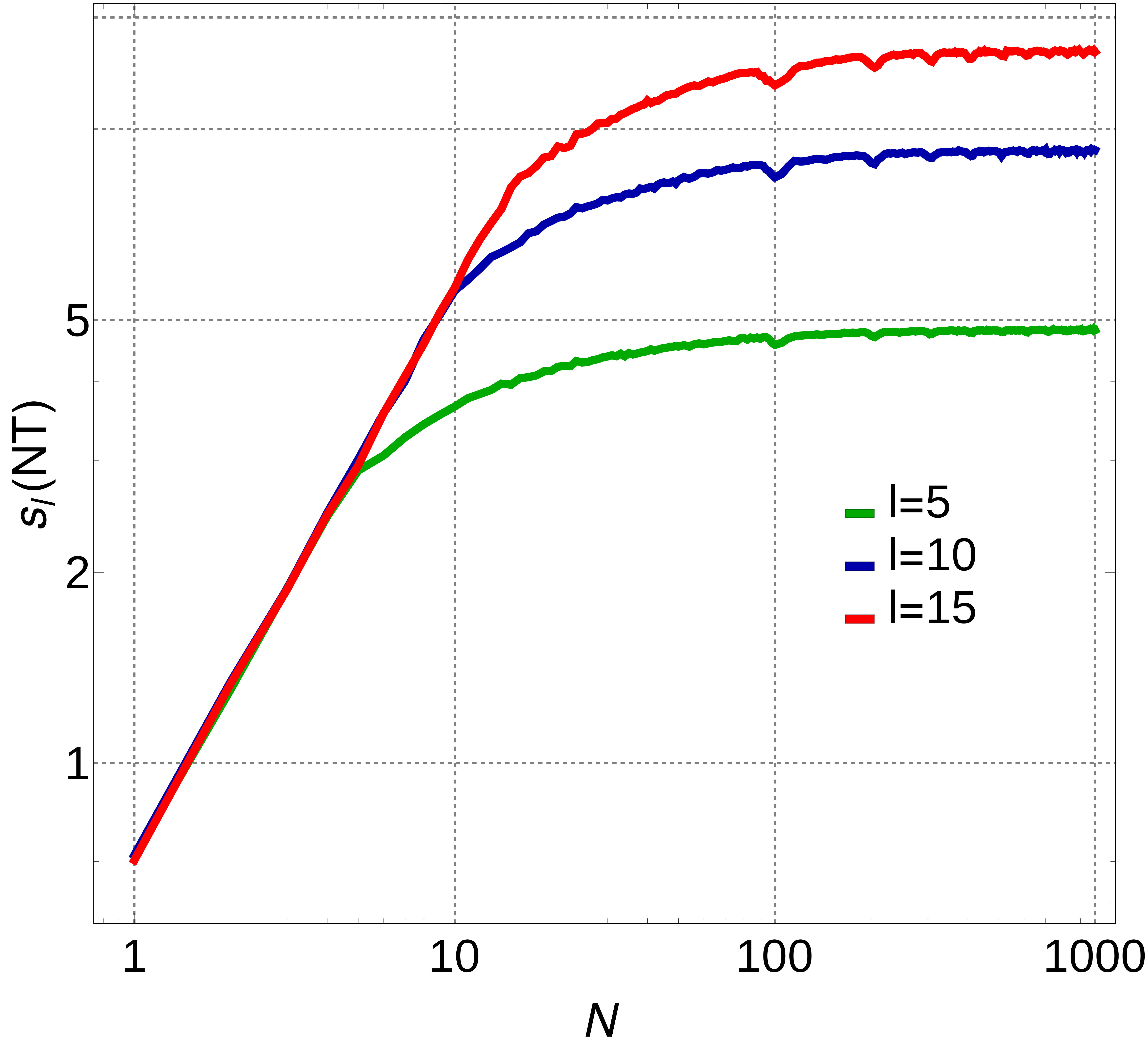}
			\label{fig_deevn}}
		\caption{ (Color online) {(a) The entanglement entropy for the TFIM as function of stroboscopic intervals with different values of subsystem size $l$ for perfectly periodic drive ($p=1$). (b) Configuration averaged EE plotted as function of stroboscopic intervals for the aperiodic drive ($p=0.5$). We have chosen sinusoidal driving with frequency $\omega=10$, amplitude $\alpha=5$, total system size $L=100$ in both the cases and for the aperiodic case the EE is averaged over 1000 configurations. Note that we use logarithmic scale in both the axes to show the initial linear growth as well as the saturation in large time limit. In both the cases a linear growth of EE is present up to a crossover time $t^*=l/2$ and the quasi-revivals occur at $t_r=L/v_{max}$, which is independent of $l$. See the gridlines along y-axis to note the saturation value.}}
		
\end{figure*}

We first consider a sudden quench at $t=0$ from the unentangled state at $h^0=\infty$  to the critical point $h=1$; following this, the spin chain is driven with the protocol given in Eq.~\eqref{eq_protocol} as, $h(t)= 1 + \gamma (t)$  with $f(t)=\alpha \sin (\omega t)$.  For the perfectly periodic situation $p=1$, the stroboscopic evolution is governed by the Floquet Hamiltonian.
% in the Section~I. 
Constructing  the  Floquet operator  $\bold{\cal{F}}_k(T)$ for each $k$ mode by solving the Bogoliubov-de Gennes equation, the state at time $t=NT$ can be easily obtained from the product of $N$ Floquet evolution operators acting on the initial state:
\begin{eqnarray}
\lvert\psi_k(NT) \rangle &=&  \left[\bold{\cal{F}}_k(T)\right]^N \lvert\psi_k(0) \rangle\\ \nonumber
&=&  \left( v_k(NT) + u_k(NT)c_k^{\dagger}c_{-k}^{\dagger}\right)\lvert0\rangle
\end{eqnarray}  
Having obtained the state $\ket{\psi(NT)}=\prod_k \lvert\psi_k(NT) \rangle$, one can calculate the following correlation matrix elements
\begin{eqnarray}
	C_{mn} &=& \frac{1}{L} \sum_k \lvert u_k(NT) \rvert^2 e^{-ik(m-n)}, \\
	F_{mn} &=& \frac{1}{L} \sum_k  u_k^*(NT) v_k(NT) e^{-ik(m-n)}.
\end{eqnarray}
The  knowledge of the correlation matrix $\mathcal{C}_l(NT)$ enables us to calculate $S_l(NT)$ using the Eq.~\eqref{eq_vnecormat}.

For perfectly periodic drive, it can be shown  that growth of the EE is linear in  time  up to a crossover time $t^*\sim l/2$ and it saturates to the periodic steady state value proportional to the subsystem size $l$ in the asymptotic limit. Remarkably although this steady value can be obtained  from  the diagonal ensemble described by the density matrix
 $$\rho_k^D(p=1) = \left(\begin{array}{cc} \lvert r_k^+\rvert^2 & 0 \\ 0 & \lvert r_k^-\rvert^2 \\ \end{array}\right),$$
it does not saturate to the  thermal value  in the asymptotic limit (see Fig.~\ref{fig_peevn}). The initial linear growth  is dictated by the maximum stroboscopic group velocity of the Floquet quasi-particles, which in turn is  given by the momentum derivative of the dispersion relations of Floquet spectrum as $v_{max}=\lvert (d\mu_k/dk) \rvert_{max}$. The linear
rise of the EE  can be physically understood in terms of the excitations that occur at each point of the system in terms of a pair of left and right moving highly entangled quasi-particles with a constant velocity $v\leq v_{max}$. Now the amount of entanglement between the subsystem and the rest of the system is given by the number of pairs of entangled quasi-particles that have one particle in the subsystem while the other one is outside. So the EE increases linearly in time up to a time $t^*=l/(2 v_{max})$, when the quasi-particle originating at the middle of the subsystem reaches the boundary. After the crossover time $t^*$, the EE dos not abruptly  saturate to its asymptotic value  due to existence of  the slow moving quasi-particles. Finally when quasi-particle pairs from every point of the subsystem is outside, EE saturates to a value which is proportional to $l$. We further note that  the quasi-revivals observed in Fig.~\ref{fig_peevn} occur due to the finite size effects. These quasi-revivals first occur at time $t_r=L/v_{max}$, when the pair of quasi-particles with velocity $v_{max}$ again travel back to the same point in the subsystem.

We shall now analyse the growth of the EE  for the aperiodic driving ($0 < p < 1$) protocol.  We calculate the EE using a method identical to that discussed above for
the $p=1$ case for a particular disorder configuration: 
%we can still write a BCS like form of the state (Eq.~\eqref{eq_genu})  $t=NT$ for a particular configuration among $N$ number of $\bold{U}(g_n)$'s and calculate the EE $(S_l(N))$ for that particular configuration. But at the end, since we have to 
take the  average over several configurations to get the configuration  averaged value of the  EE $(\overline{S_l(N)})$.
%For any finite value of $N$, although we can calculate %averaged value of all the  correlation matrix elements, %still we can't use our formalism of Floquet coin toss %dynamics to calculate $\overline{S_l(N)}$. This is %because of the configuration average of the Von Neumann %entropy of correlation matrix is not the same as Von %Neumann entropy of the configuration averaged %correlation matrix.
The result thus obtained  is shown in Fig.~\ref{fig_deevn}. Remarkably, the dynamical features such as short time linear growth, quasi-revivals do indeed persist  even when
the dynamics is aperiodic. Further, the maximum speed of the propagation of quasi-particles remains nearly unaltered for $p \neq 1$ as both the $t^*$ and $t_r$ are remain same as the $p=1$ case. However comparing Figs.~\ref{fig_peevn}
and \ref{fig_deevn}, we note sharp contrast: In the aperiodic case, the  EE increases with time till it saturated to the maximum possible value which is the thermal value $s_l(\infty) = l$.
Therefore a temporal disorder in the periodic driving not only leads to a diagonal ensemble but remarkably renders an infinite temperature ensemble given by the density matrix 
$$\rho_{D}(p\neq 1) = \left(\begin{array}{cc} \frac 1 {2} & 0 \\ 0 & \frac 1 {2} \\ \end{array}\right).$$
  
We can use the disorder matrix formalism  to analytically establish that the  averaged asymptotic value of entanglement entropy $\overline{S_l(\infty)}$ is indeed $l$. 
%We know that for a ergodic system long time averaged value of some observable is equal to the ensemble averaged value. 
Due to underlying ergodicity,  in the limit $N\rightarrow\infty$, the  EE calculated for a particular configuration happens to be the same as the configuration averaged. Within the $D$-matrix
formalism, as elaborated  in  the Appendix \ref{asymptotic_ee}, one can show  that the averaged correlations $\overline{\langle c_{k}^{\dagger}c_{k}\rangle} $ and $\overline{\langle c_{k}^{\dagger} c_k^{\dagger} \rangle }$ become $1/2$ and $0$ respectively in the limit $N\to \infty$, which in turn,  leads a simple form of the correlation matrix with the elements,
\begin{eqnarray}
\overline{C_{mn}} &=&   \frac{1}{2L} \sum_k   e^{-i k (m-n)}=\frac{\delta_{mn}}{2}, \\ \nonumber
\overline{F_{mn}} &=&  0.
\end{eqnarray}
% This is exactly equal to $C_{mn}(\infty)$ and $F_{mn}(\infty)$ i.e. for a particular configuration with $N\to \infty$. 
 We therefore have a diagonal form of $2l\times 2l$ correlation matrix $\mathcal{C}_l(\infty)$ with all the $2l$ diagonal elements being equal to $1/2$. This corresponds to maximum value of entanglement entropy $S_l(\infty)=l\left(-\log_2[1/2]\right)=l$, which is essentially the thermal (corresponding to an infinite temperature ensemble) value of the EE.

\section{Conclusions}
In the present work, we have studied the aperiodic driving of two Jordan-Wigner solvable models. Considering the HCB chain, we have shown that for
minimal aperiodicity in the driving the residual energy asymptotically reaches the value corresponding to the infinite temperature ensemble while in the perfectly
periodic case ($p=1$), the residual energy saturates to a periodic steady state value. We then proceed to
study the fate of the current present in the initial state (generated due to the application of a twist) as a consequence of the aperiodic driving. In the $p=1$
situation, the initial  current vanishes asymptotically in the limit of $\omega \to \infty$ for $\delta$-kicking leading to the so-called dynamical localisation.
In the case of sinusoidal driving, there is no such localisation in the large $\omega$ limit, rather the maximum stroboscopic group velocity attains
a constant value. Interestingly, we have shown that  with the onset of aperiodicity, the current vanishes in the asymptotic limit for both the protocols for any frequency and any $p \neq 1$. Using
the $D$-matrix formalism, we have shown this vanishing of current within an exact analytical framework.

Regarding the entanglement entropy, we have established that for any $p\neq 1$ and any frequency, the system is described by a thermal diagonal ensemble,
i.e., interestingly $S_l$ attains the thermal value of $l$ in the asymptotic limit unlike the $p=1$ situation when the resulting diagonal ensemble in not thermal.
 Using the $N \to \infty$ structure of the $D$-matrix, we have established that 
indeed the $S_l$ attains the thermal  value $l$ in that limit. Remarkably, short-time linear growth of $S_l$ and quasi-revival structure remain robust
against the aperiodic perturbation and persists irrespective of the value of $p$, which implies the maximum stroboscopic group velocity is unaffected for aperiodic driving.
This linear growth and the asymptotic saturation to the thermal value is similar to the temporal growth of  $S_l$ as observed following a sudden quench of a non-integrable model \ct{huse13}; this implies that the temporal aperiodicity  in periodic driving breaks the underlying integrability of the model.

\begin{acknowledgments}
We acknowledge  Souvik Bandyopadhyay, Sourav Bhattacharjee, Sudarshana Laha and Sourav Nandy for fruitful discussions. AD acknowledges SERB, DST for financial support.
\end{acknowledgments}

\appendix 

\section{Construction of the $D$-matrix  and its structure in the limit $N \to \infty$}\label{app_dmat}
In this appendix, we will show the emergence of the $4\times4$ disorder matrix for each momentum mode $k$ as result of the binary disorder present on top of the periodic driving and it's structure in the assymptotic limit of $N$.

Recalling the equation \eqref{eq_expt}, under the application of coin-toss like aperiodic drive, the expectation value of some operator $\hat{O}$ after $N$ stroboscopic intervals is given by,
\begin{eqnarray} 
\langle \hat{O}(NT) \rangle & = & \langle \psi(0) \rvert \bold{U}^{\dagger}(g_1) \bold{U}^{\dagger}(g_2) .......  \bold{U}^{\dagger}(g_N) \nonumber \\
&\times& \hat{O} \bold{U}(g_N)........\bold{U}(g_2) \bold{U}(g_1) \lvert \psi(0) \rangle.
\label{eq_qavg}
\end{eqnarray}
Introducing $2(N+1)$ number of identity operators $\sum_{j_n =1}^{2} |j_n\rangle\langle j_n|$  in the basis of the intial Hamiltonian $H_k^0$, we can write the above equation as,
\begin{widetext}
\begin{eqnarray}
\langle \hat{O}(NT) \rangle &=& \sum_{\substack{ j_0,j_2,....,j_N \\i_0, i_1, ....,i_N}} \langle \psi(0) \rvert j_0 \rangle \langle j_0 \rvert \bold{U}^{\dagger}(g_1) \lvert j_1 \rangle \langle j_1 \rvert \bold{U}^{\dagger}(g_2) \lvert j_2 \rangle \dots \langle j_{N-1} \rvert \bold{U}^{\dagger}(g_N) \lvert j_N \rangle\langle j_N \rvert \hat{O} \lvert i_N \rangle \\ \nonumber
&& \hspace{2.0cm} \times  \langle i_N \rvert \bold{U}(g_N) \lvert i_{N-1} \rangle \dots \langle i_2 \rvert \bold{U}(g_2) \lvert i_1 \rangle \langle i_1 \rvert  \bold{U}(g_1) \lvert i_0 \rangle \langle i_0 \lvert \psi(0) \rangle  \\\nonumber
&=& \sum_{\substack{ j_0,j_2,....,j_N \\i_0, i_1, ....,i_N}} \langle \psi (0) \rvert j_0 \rangle \langle j_N \rvert H_0 \lvert i_N \rangle \langle i_0 \lvert \psi (0) \rangle \left( \prod_{m=1}^{N} \langle j_{m-1} \rvert \bold{U}^{\dagger}(g_m) \lvert j_{m} \rangle \langle i_{m} \rvert \bold{U}(g_m) \lvert i_{m-1} \rangle \right)
\end{eqnarray}
As the random binary variables $g_m$ ( with probability $P(g_m)$) are uncorrelated, we can evaluate the disorder average of the avobe quantity as,	
\begin{eqnarray} \nonumber
\overline{\langle \hat{O}(NT) \rangle} &=&  \sum_{\substack{ j_0,j_2,....,j_N \\i_0, i_1, ....,i_N}} \langle \psi (0) \rvert j_0 \rangle \langle j_N \rvert H_0 \lvert i_N \rangle \langle i_0 \lvert \psi (0) \rangle 
\left[ \prod_{m=1}^{N} \left( \sum_{g_m = 1, 0} P(g_m) \langle j_{m-1} \rvert \bold{U}^{\dagger}(g_m) \lvert j_{m} \rangle \langle i_{m} \rvert \bold{U}(g_m) \lvert i_{m-1} \rangle \right) \right] \\\nonumber
%& =&  \sum_{\substack{ j_0,j_2,....,j_N \\i_0, i_1, ....,i_N}}  \langle \Psi(0) \rvert j_0 \rangle \langle j_N \rvert H_0 \lvert i_N \rangle \langle i_0 \lvert \Psi(0) \rangle \\\nonumber
%&& \hspace{1.0cm} \times \left[ \prod_{m=1}^{N}  \left(  e^{iT\left(\epsilon_{j_{m-1}}-\epsilon_{i_{m-1}}\right)} p \delta_{j_{m-1},j_{m}} \delta_{i_{m},i_{m-1}} + %(1-p)   \langle j_{m-1} \rvert U_0^{\dagger} \lvert j_{m} \rangle \langle i_{m} \rvert U_0 \lvert i_{m-1} \rangle  \right) \right] \\ \nonumber
&=& \sum_{j_0,j_N,i_0,i_N}^{} \langle \psi(0)\rvert j_0 \rangle \langle j_N \rvert \hat{O} \lvert i_N \rangle \langle i_0 \rvert \psi(0) \rangle  \Big[ {D}^N \Big]_{j_0 j_N i_0 i_N}
\end{eqnarray}
where in the right hand side of the equation $D^N$ is the $N^{th}$ power of a $4\times4$ matrix $D$ with the following matrix elements:
\begin{equation}
D_{j_1,j_2,i_1,i_2} \equiv \left((1-p) e^{iT\left(E_{j_{1}}-E_{i_{1}}\right)} \delta_{j_1,j_2} \delta_{i_1,i_2} + p \langle j_2 \rvert \boldmath{\cal{F}}^{\dagger}(T) \lvert j_1 \rangle \langle i_1 \rvert \boldmath{\cal{F}}(T) \lvert i_2 \rangle \right),
\label{eq_dmat_app}
\end{equation}

\end{widetext}

As discussed in the main text, both the HCB chain and the one dimensional TFIM can be decoupled into two level systems for each momentum mode $k$. We shall henceforth, consider the $D$-matrix for a particular
momentum mode $k$;
to investigate the structure of the $D(k)$-matrix in the asymptotic limit $N\rightarrow\infty$, we shall  consider a general form of the Floquet evolution operator $\mathcal{F}_k(T)$ in the basis of the eigenstates of the initial Hamiltonian $H_k^0$ as follows,
\begin{equation}
\mathcal{F}_k(T) \doteq \left(
\begin{array}{cc}
F_{11}(k) & F_{12}(k) \\
-F_{12}^*(K) & F_{11}^*(k) \\
\end{array}
\right)
\label{eq_matrix2by2}
\end{equation}
where we have $\left|F_{11}(k)\right|^2+\left|F_{12}(k)\right|^2 =1$. So the general form of the $D$-matrix ( see Eq.~\eqref{eq_dmat_app}) has the following structure,
\begin{widetext}
	\begin{equation}
		D_k = \left(
		\begin{array}{cccc}
		(1-p)+ p \left| F_{11}(k)\right|^2 & p F_{11}^*(k)F_{12}(k) & p F_{12}^* (k)F_{11}(k) & p \left|F_{12}(k)\right|^2 \\\\
		-p F_{11}^*(k)F_{12}^*(k) & (1-p)\exp{[-i\Delta \phi_k T]}+p F_{11}^*(k)F_{11}^*(k) & -p F_{12}^*(k)F_{12}^*(k) & p F_{12}^*(k) F_{11}^*(k) \\\\
		-p F_{12}(k) F_{11}(k) & -p F_{12}(k)F_{12}(k) & p\exp{[i\Delta\phi_k T]}+ p F_{11}(k) F_{11}(k) & p F_{11}(k) F_{12}(k) \\\\
		p \left| F_{12}(k)\right|^2  & -p F_{12}(k) F_{11}^*(k) & -p F_{11}(k) F_{12}^*(k) & (1-p) +p \left| F_{11}(k)\right|^2 \\
		\end{array}
		\right) \\
		\label{eq_D_matrix}
	\end{equation}
\end{widetext}
where $\Delta \phi_k = E_2(k)-E_1(k)$, the energy gap of $H_k^0$.  To investigate the structure of $D$-matrix in the asymptotic limit, we analyze its eigenvalues. To compute four eigenvalues, one needs to solve the following eigenvalue equation:
\begin{equation}
\left[\lambda (k)-s(k)\right]f(\lambda ,k)=0
\end{equation}
where $s( k) =\left\{p+(1-p)\left( \left| W_{11}(k)\right|^2+\left| W_{12}(k)\right|^2\right)\right\}=1$ as $\left( \left| W_{11}(k)\right|^2+\left| W_{12}(k)\right|^2\right)=1$,
and $f(k,\lambda)$ is  a third degree polynomial with all real coefficients.
Thus, it is obvious that always one of the eigenvalues $\lambda_1 =1$ with normalized eigenvector, $v_1=\frac{1}{\sqrt{2}} \left(1, 0, 0, 1 \right)^T$ for each $k$ mode. 
While one eigenvalue always sticks to unity, we have seen that other eigenvalues (one real and the other two complex conjugates of each other) will have a value (or modulus) less
than unity due to the presence of off-diagonal terms in matrix in \eqref{eq_D_matrix} and  vanish in $D^N(k)$ when $N \to \infty$. Given the simple structure of the diagonal form of the $\lim_{N\to\infty}D^N(k) = diag(1,0,0,0)$, it is easy to verify the following form of the $D$-matrix in $N\to \infty$ limit,
\begin{equation}
\lim_{N\to\infty}D^N(k)=\frac{1}{2}\left(
\begin{array}{cccc}
1 & 0 & 0 & 1 \\
0 & 0 & 0 & 0 \\
0 & 0 & 0 & 0 \\
1 & 0 & 0 & 1 \\
\label{eq_dninf_app}
\end{array}
\right)
\end{equation}

It is noteworthy that the asymptotic form of the $D$-matrix  given above is independent of the protocol (i.e., identical for the $\delta$-kick and the sinusoidal driving), its frequency and amplitude and the bias of coin-toss $p$. Importantly, while the eigenvalue $\lambda=1$ dectate the universal asymptotic behavior, the other eigenvalues having value or modulus less than unity determind all the non-universal early time behavior of the quantities of interest and depend on the bias $p$, amplitude $\alpha$ and the frequency $\omega$.

\section{Entanglement entropy in the asymptotic limit $N \rightarrow \infty $ }\label{asymptotic_ee}
In this section, we will calculate the asymptotic value of entanglement entropy using the disorder matrix formalism for Floquet coin toss dynamics. To calculate $\overline{S_l(\infty)}$, we use the Eq.~\eqref{eq_davg} for each $k$ mode, where the operator $\hat{O}$ replaced by $c_{k_1}^{\dagger} c_{k_2}$ and $c_{k_1}^{\dagger} c_{k_2}^{\dagger}$. Here we used the eigenstates of $H_k^0$ as basis,
\begin{eqnarray}
	\lvert 1 \rangle \equiv \lvert \psi_k^g(0) \rangle &=& \left( v_k^0 + u_k^0 c_k^{\dagger}c_{-k}^{\dagger}\right)\lvert0\rangle \\
	\lvert 2 \rangle \equiv \lvert \psi_k^e(0) \rangle  &=& \left( u_k^{0 *} - v_k^{0 *} c_k^{\dagger}c_{-k}^{\dagger}\right)\lvert0\rangle
\end{eqnarray}

In this basis, one can obtain the following correlations of the momentum modes,
\begin{eqnarray}
\langle 1 \rvert c_{k_1}^{\dagger}c_{k_2} \lvert 1 \rangle & = & \lvert u_k^0 \rvert^2 \delta_{k_1,k_2} \\
\langle 2 \rvert c_{k_1}^{\dagger}c_{k_2} \lvert 2 \rangle & = & \lvert v_k^0 \rvert^2 \delta_{k_1,k_2} \\
\langle 1 \rvert c_{k_1}^{\dagger}c_{k_2} \lvert 2 \rangle & = & - u_k^{0*} v_k^{0*} \delta_{k_1,k_2} \\
\langle 2 \rvert c_{k_1}^{\dagger}c_{k_2} \lvert 1 \rangle & = & - u_k^{0} v_k^{0} \delta_{k_1,k_2}
\end{eqnarray}

\begin{eqnarray}
\langle 1 \rvert c_{k_1}^{\dagger}c_{k_2}^{\dagger} \lvert 1 \rangle & = & u_k^{0*}v_k^0 \delta_{k_1,-k_2} \\
\langle 2 \rvert c_{k_1}^{\dagger}c_{k_2}^{\dagger} \lvert 2 \rangle & = & -u_k^{0*} v_k^0 \delta_{k_1,-k_2} \\
\langle 1 \rvert c_{k_1}^{\dagger}c_{k_2}^{\dagger} \lvert 2 \rangle & = & u_k^{0*} u_k^{0*} \delta_{k_1,-k_2} \\
\langle 2 \rvert c_{k_1}^{\dagger}c_{k_2}^{\dagger} \lvert 1 \rangle & = & v_k^{0} v_k^{0} \delta_{k_1,-k_2}
\end{eqnarray}

\begin{widetext}
	
So the averaged correlations after $N$ number of stroboscopic intervals are given by,
	\begin{eqnarray}\label{eq_avgck}
	\overline{\langle c_{k_1}^{\dagger}c_{k_2} \rangle} &=& \left( \lvert u_k^0 \rvert^2\left[D^N\right]_{1111} +\lvert v_k^0 \rvert^2\left[D^N\right]_{2121} -u_k^{0*} v_k^{0*}\left[D^N\right]_{1121} - u_k^{0} v_k^{0}\left[D^N\right]_{2111} \right)\delta_{k_1,k_2=k} \\ \nonumber
	\overline{\langle c_{k_1}^{\dagger}c_{k_2}^{\dagger} \rangle} &=& \left( u_k^{0*} v_k^0\left[D^N\right]_{1111} - u_k^{0*} v_k^0 \left[D^N\right]_{2121} + u_k^{0 *} u_k^{0*}\left[D^N\right]_{1121} - v_k^{0} v_k^{0}\left[D^N\right]_{2111} \right)\delta_{k_1,-k_2=k}
	\end{eqnarray}
	
\end{widetext}

The averaged value of the elements of correlation matrices $C$ and $F$ can be calculated by Fourier transforming the above correlations, 
\begin{equation}
	\overline{C_{mn}} =   \frac{1}{L} \sum_k \overline{\langle c_{k}^{\dagger}c_{k} \rangle} e^{-i k (m-n)} 
\end{equation}
and
\begin{equation}
\overline{F_{mn}} =   \frac{1}{L} \sum_k \overline{\langle c_{k}^{\dagger} c_k^{\dagger} \rangle } e^{-i k (m-n)}
\end{equation}

Using the form of the $D$-matrix in the $n\to \infty$ limit (Eq.~\eqref{eq_dninf}), the above quantities take the following simple form,
\begin{eqnarray}
	\overline{C_{mn}} &=&   \frac{1}{2L} \sum_k   e^{-i k (m-n)}=\frac{\delta_{mn}}{2} \\ \nonumber
	\overline{F_{mn}} &=&  0
\end{eqnarray}

as from the Eq.~\eqref{eq_avgck} $\overline{\langle c_{k}^{\dagger}c_{k}\rangle} $ and $\overline{\langle c_{k}^{\dagger} c_k^{\dagger} \rangle }$ become $1/2$ and $0$ respectively. This is exactly equal to $C_{mn}(\infty)$ and $F_{mn}(\infty)$ i.e. for a particular configuration with $N\to \infty$. So we have a diagonal form of $2l\times 2l$ correlation matrix $\mathcal{C}_l(\infty)$ with all the $2l$ diagonal elements being equal to $1/2$.

\end{document}